\documentclass[11pt,letterpaper]{article}
\pdfoutput=1
\usepackage{jcappub}

\usepackage{appendix}
\usepackage{array}
\newcolumntype{x}[1]{>{\centering\arraybackslash}p{#1}}

\usepackage{epsfig}
\usepackage{graphicx}
\usepackage{dcolumn}
\usepackage{amsmath}
\usepackage{enumerate}

\usepackage{color}
\usepackage{ulem}
\hypersetup{colorlinks=false}

\newcommand{\eg}{e.g.~}
\newcommand{\ie}{i.e.~}

\newcommand{\beq}{\begin{equation}}
\newcommand{\eeq}{\end{equation}}
\newcommand{\ud}{\text{d}}

\newcommand{\Eq}[1]{eq.~\eqref{#1}}
\newcommand{\mDM}{m}
\newcommand{\ER}{E_\text{R}}
\newcommand{\Ed}{E'}
\newcommand{\vesc}{v_\text{esc}}
\newcommand{\vmin}{v_\text{min}}

\newcommand{\bfv}{\mathbf{v}}

\newcommand{\eR}{\mathcal{R}}

\title{Update on Light WIMP Limits: \\ LUX, lite and Light}

\author[a]{Eugenio Del Nobile,}
\author[a]{Graciela B. Gelmini,}
\author[b]{Paolo Gondolo,}
\author[a]{and Ji-Haeng Huh}

\affiliation[a]{Department of Physics and Astronomy, UCLA,\\
475 Portola Plaza, Los Angeles, CA 90095, USA}
\affiliation[b]{Department of Physics and Astronomy, University of Utah,\\
115 South 1400 East \#201, Salt Lake City, UT 84112, USA}

\emailAdd{delnobile@physics.ucla.edu}
\emailAdd{gelmini@physics.ucla.edu}
\emailAdd{paolo@physics.utah.edu}
\emailAdd{jhhuh@physics.ucla.edu}

\abstract{We reexamine the current direct dark matter data including the recent CDMSlite and LUX data, assuming that the dark matter consists of light WIMPs, with mass close to 10 GeV/$c^2$ with spin-independent and isospin-conserving or isospin-violating interactions. We compare the data with a standard model for the dark halo of our galaxy and also in a halo-independent manner. In our standard-halo analysis, we find that for isospin-conserving couplings, CDMSlite and LUX together exclude the DAMA, CoGeNT, CDMS-II-Si, and CRESST-II possible WIMP signal regions. For isospin-violating couplings instead, we find that a substantial portion of the CDMS-II-Si region is compatible with all exclusion limits.  In our halo-independent analysis, we find that for isospin-conserving couplings, the situation is of strong tension between the positive and negative results, as it was before the LUX and CDMSlite bounds, which turn out to exclude the same possible WIMP signals as previous limits. For isospin-violating couplings, we find that LUX and CDMS-II-Si bounds together exclude or severely constrain the DAMA, CoGeNT and CRESST-II possible WIMP signals.}


\begin{document}

\maketitle

\section{Introduction}
The presence of dark matter (DM) in the universe is now an established fact, that has been confirmed once more by the recent precise measurements of the Planck satellite~\cite{Ade:2013zuv}. Many different particle candidates exist as possible explanations for the DM. A particular class of candidates, the WIMPs, (for weakly interacting massive particles), is very actively searched for. WIMPs are particles with weakly interacting cross sections and masses in the 1 GeV/$c^2$ -- 10 TeV/$c^2$ range.

Of particular interest are light WIMPs, with mass around 1 -- 10 GeV/$c^2$, because at present four direct dark matter search experiments (DAMA~\cite{Bernabei:2010mq}, CoGeNT~\cite{Aalseth:2010vx, Aalseth:2011wp, Aalseth:2012if,  Aalseth:2014eft, Aalseth:2014jpa}, CRESST-II~\cite{Angloher:2011uu}, and more recently CDMS-II-Si~\cite{Agnese:2013rvf}), have data that may be interpreted as signals from DM particles in the light WIMPs range. DAMA~\cite{Bernabei:2010mq} and CoGeNT~\cite{Aalseth:2011wp,  Aalseth:2014eft, Aalseth:2014jpa} report annual modulations in their event rates, compatible with those expected for a DM
 signal~\cite{Drukier:1986tm}. CoGeNT~\cite{Aalseth:2010vx, Aalseth:2012if,  Aalseth:2014jpa}, CRESST-II~\cite{Angloher:2011uu}, and very recently CDMS-II-Si ~\cite{Agnese:2013rvf}, observe an excess of events above their expected backgrounds; these excesses may be interpreted as due to DM WIMPs.

However, other experiments do not observe significant excesses above their estimated background, thus setting upper limits on DM WIMPs. The most stringent limits on the average (unmodulated) rate for light WIMPs were set until recently by the XENON10~\cite{Angle:2011th}, XENON100~\cite{Aprile:2011hi, Aprile:2012nq}, and CDMS-II-Ge~\cite{Ahmed:2010wy} experiments, with the addition of SIMPLE~\cite{Felizardo:2011uw}, PICASSO~\cite{Archambault:2012pm} and COUPP~\cite{Behnke:2012ys} for spin-dependent and isospin-violating interactions. CDMS-II-Ge~\cite{Ahmed:2012vq} also constrains directly the amplitude of an annually modulated signal.

Recently, the CDMS collaboration (now in the SuperCDMS phase operating at the Soudan Underground Laboratory) published the results of a very low threshold run using a single germanium detector~\cite{Agnese:2013lua}. Operated in a new mode that makes use of a relatively high bias voltage across the detector, the experiment is called CDMSlite (for CDMS Low Ionization Threshold Experiment) and yields very high sensitivity to light WIMPs. With a small exposure of 6.3 kg $\times$ days, the collaboration was able to constrain new WIMP-nucleon spin-independent (SI) parameter space for WIMP masses below 6 GeV/$c^2$.

Most recently the first results from the LUX DM experiment were presented~\cite{Akerib:2013tjd}. The Large Underground Xenon (LUX) experiment, like the XENON experiment, is a dual-phase xenon time-projection chamber. It is operated at the Sanford Underground Research Facility. With an exposure of 85.3 live-days $\times$ 118.3 kg of fiducial volume, the LUX collaboration set bounds on WIMPs with standard-halo SI interactions more restrictive than previously existing bounds in a large region of the WIMP mass--DM-proton cross section $m$--$\sigma_p$ parameter space.

Here we derive the bounds due to CDMSlite, LUX and other experimental data, for SI interactions with  both  isospin-conserving and isospin-violating~\cite{Kurylov:2003ra, Feng:2011vu} interactions. We compare all bounds and regions for light WIMPs both in the $m$--$\sigma_p$ parameter space, assuming the Standard Halo Model (SHM), and in a halo model-independent manner~\cite{Fox:2010bz, Frandsen:2011gi, Gondolo:2012rs, Frandsen:2013cna, DelNobile:2013cta, DelNobile:2013cva} (see also~\cite{HerreroGarcia:2011aa, HerreroGarcia:2012fu, Bozorgnia:2013hsa}).

\section{Data analysis}
\label{data}

We consider data from DAMA~\cite{Bernabei:2010mq}, CoGeNT~\cite{CoGeNT2011release, Aalseth:2012if, Aalseth:2014eft}, CRESST-II~\cite{Angloher:2011uu}, CDMS-II-Ge low threshold analysis \cite{Ahmed:2010wy}, CDMS-II-Ge annual modulation analysis~\cite{Ahmed:2012vq}, CDMS-II-Si~\cite{Agnese:2013rvf}, XENON10 S2-only analysis~\cite{Angle:2011th}, XENON100~\cite{Aprile:2012nq}, and SIMPLE~\cite{Felizardo:2011uw}. The analysis we perform on these data is described in~\cite{DelNobile:2013cta, DelNobile:2013cva}. The CRESST-II~\cite{Angloher:2011uu} and the
very recent 2014  CoGeNT modulation~\cite{Aalseth:2014eft} data are analyzed as explained below only for our halo-independent analysis, but the  regions (and in the case of CoGeNT an upper limit too)  shown in our  SHM  based plots are those presented  by the respective collaborations in Refs.~\cite{Angloher:2011uu} and~\cite{Aalseth:2014jpa} (we do analyze the older CoGeNT 2011-2012 data~\cite{CoGeNT2011release, Aalseth:2012if} in all cases). We corrected a mistake in the background subtraction of CRESST-II in our previous halo-independent analysis, which resulted in a change in the lowest speed data point which, however, does not change qualitatively our conclusions. 

In addition, we consider here the recent results from CDMSlite~\cite{Agnese:2013lua} and LUX~\cite{Akerib:2013tjd}. For CDMSlite, we read off the recoil energy spectrum data from Fig.~1 of Ref.~\cite{Agnese:2013lua}, using the histogram with 10 eVee bins for measured energies between 0.10 and 1.60 keVee, and the histogram with 75 eVee bins above 1.60 keVee. To obtain the CDMSlite limits we used the Maximum Gap Method~\cite{Yellin:2002xd}, which requires unbinned data. We construct these data by dividing each bin with multiple events into enough bins of equal width so that in the end we have one event per bin; we then assign to each event the middle energy of the bin containing it. Following Ref.~\cite{Agnese:2013lua} we only consider the energy range 0.17 to 7.00 keVee, take the efficiency to be 98.5\% and the energy resolution to be $\sigma = 14$ eVee. To be consistent with the analysis of the CoGeNT data we use the same quenching factor used by CoGeNT, $Q_\text{Ge} = 0.2 \ER^{0.12}$ where $\ER$ is the nuclear recoil energy (modifications of the quenching factor affect the bound as illustrated in Fig.~4 of Ref.~\cite{Agnese:2013lua}).

Regarding the calculation of upper limits from LUX, we do not have access to all the information needed to recompute the bound in Ref.~\cite{Akerib:2013tjd} or extend it to a halo-independent analysis (or to other kinds of WIMP-nucleus interactions). The major obstacle encountered is the overlap between the background and expected signal distributions in the S1--$\log_{10}$(S2/S1) plane.\footnote{Looking at Fig.~4 of Ref.~\cite{Akerib:2013tjd}, one might think that the nuclear recoil distribution (red band) does not overlap with the  electron recoil distribution (blue band). This thinking is, however, deceptive for two reasons. First, the dashed blue and red contours in Figs.~3 and 4 of Ref.~\cite{Akerib:2013tjd} only show the $1.28 \sigma$ band around the mean of the distribution (solid lines), and therefore a fraction of the events are expected to leak from one band to the other. Second, the red band corresponds to the distribution of neutron calibration  events, which according to Ref.~\cite{Akerib:2013tjd}, due to systematic effects  does not coincide with the distribution expected for WIMP events.}
A region of interest in this plane is usually defined within which the number of observed events is compared with the number of expected WIMP events. Because of the just mentioned overlap in LUX, if the region of interest is large enough to include almost all of the expected signal, the number of LUX background events within it is too large to reproduce the LUX published limit without additional information. On the other hand, if the region of interest is restricted as to contain a small number of observed events, the fraction of expected signal events within the region of interest needs to be evaluated, since it enters as an efficiency in the calculation of the expected number of events, and it sensitively affects the evaluation of upper bounds on the WIMP-nucleon cross section. The fraction of expected signal events within the region of interest depends on the halo model, the WIMP model, and the S1 and S2 statistical distributions as functions of the nuclear recoil energy. We cannot reconstruct the latter.

It may be tempting to bypass the lack of information by reasoning~\cite{Gresham:2013mua} that the fraction of WIMP events below the measured mean of the neutron-recoil (NR) calibration events, \ie the solid red line in Figs.~3 and 4 of Ref.~\cite{Akerib:2013tjd}, should conservatively be one half (meaning that it should be more than 1/2 so that assuming it to be 1/2 leads to a conservative upper bound). However, the fraction of WIMP events below the NR mean line changes with the expected WIMP recoil spectrum and it cannot be guaranteed to be more than 1/2 independently of the WIMP velocity distribution or the WIMP-nucleus differential cross section. Thus choosing a signal fraction equal to 1/2 below the NR mean line cannot ensure a conservative upper limit. Choosing another constant value of the signal fraction below the NR mean line would introduce subjectiveness. We therefore proceed in the following manner.

To compute the LUX bound, we apply the Maximum Gap Method~\cite{Yellin:2002xd} to the variable S1 in the range 2--30 photoelectrons. For the observed events to use in this method, we notice that the maximum recoil energy occurring in our halo-independent analysis ($m \leqslant 9$ GeV/$c^2$, $\vmin \leqslant 10^3$ km/s) for scattering off Xe is of $\sim 12 $ keVnr.  Using the approximated recoil energy contours in Fig.~4 of Ref.~\cite{Akerib:2013tjd}, and dropping all observed events in and above the electron-recoil band (plotted at $1.28 \sigma$) in the same figure, only five observed events remain below $\sim 12$ keVnr. Thus we computed several Maximum Gap limits using either 0 observed events, or 1 event (with ${\rm S1} = 3.1$ photoelectrons), or 3 events (the previous one and those with S1 5.5 and 6.0 photoelectrons), or 5 events (the previous ones and the two with ${\rm S1} = 3.5$ photoelectrons and $\log_{10}$(S2/S1)$\simeq$2). In our SHM analysis, which is confined to $m \leqslant 30$ GeV/$c^2$ and $\vmin < 800$ km/s, we reach recoil energies well above those corresponding to ${\rm S1} = 30$ photoelectrons. In this case we also show the limit obtained by considering all the 24 observed events below the electron-recoil band in Fig.~4 of Ref.~\cite{Akerib:2013tjd}. Since our procedure does not depend on the WIMP distribution in the S1--$\log_{10}$(S2/S1) plane, and given that we use all the events below the experimentally well established electron-recoil band in Fig.~4 of Ref.~\cite{Akerib:2013tjd}, our Maximum Gap upper limits are conservative and safe to be extended to other halo models and WIMP-nucleus interactions.

In addition to the observed events we need the efficiency as a function of S1, and the distribution of S1 values for a given recoil energy. We take the S1 efficiency shown as the dashed red line in Fig.~1 of Ref.~\cite{Akerib:2013tjd} and, following the LUX analysis in Ref.~\cite{Akerib:2013tjd} and consistently with our treatment of XENON100 data, we set the counting efficiency to zero below 3.0 keVnr. We take the S1 single photoelectron resolution to be $\sigma_\text{PMT} = 0.37$ photoelectrons~\cite{Akerib:2012ys}. We reconstruct the quenching factor using the recoil energy contour lines in Fig.~4 of Ref.~\cite{Akerib:2013tjd}, assigning to each recoil energy the S1 value at the NR mean line. This procedure is approximately correct where the recoil energy contours in the figure are approximately vertical, \ie for the low S1 values on which our light WIMP limits depend the most. Even for the largest WIMP masses we reach in Fig.~\ref{m-sigma}, for which the recoil energies correspond to larger values of S1, we found that the maximum gap determining the limits are at small S1 values where the approximation is good (even when we use 24 events).

\begin{figure}[t]
\centering
\includegraphics[width=0.7\textwidth]{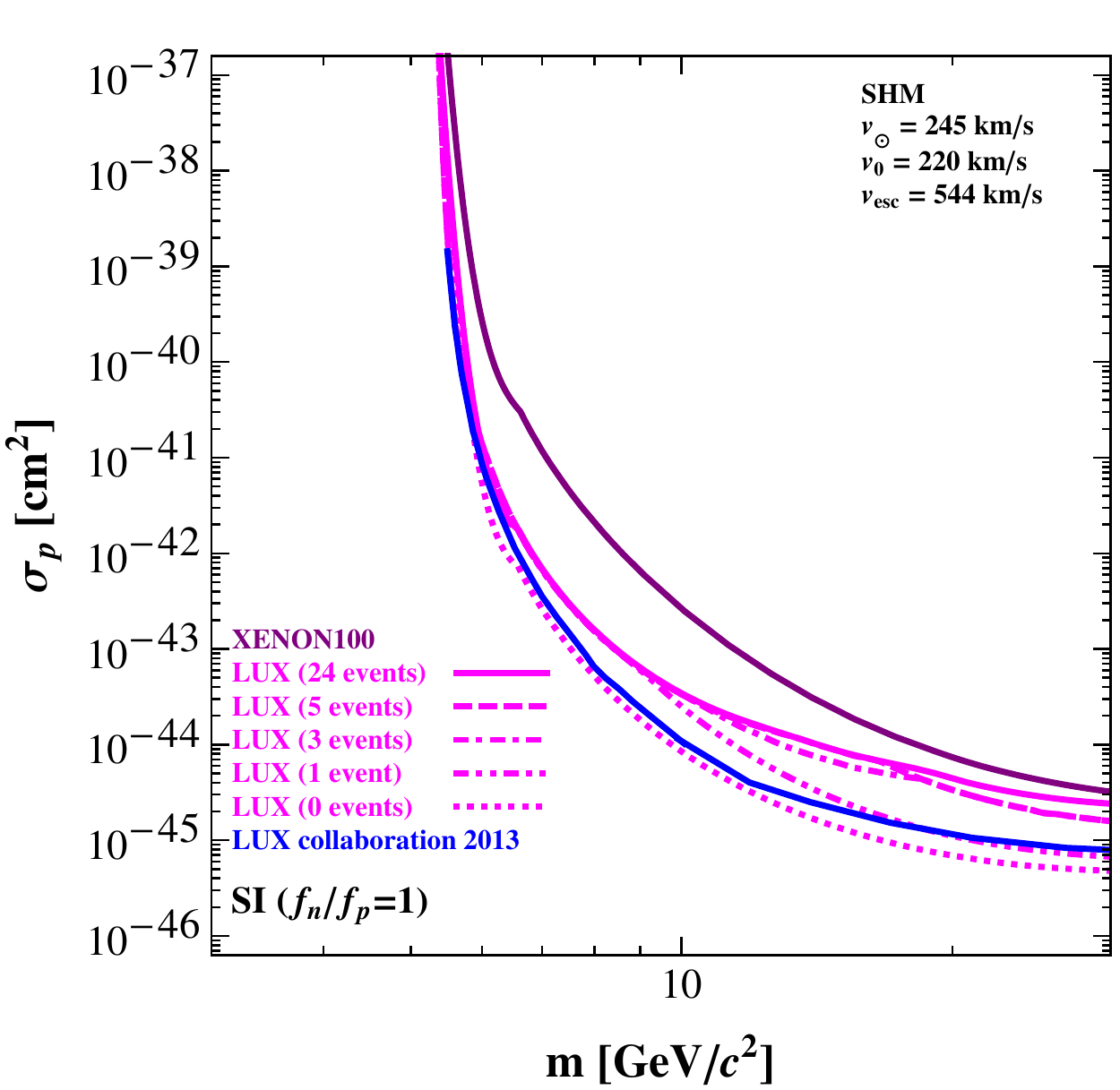}
\caption{Our $90\%$ CL LUX bounds for 0, 1, 3, 5 and 24 observed events (see Section~\ref{data}), compared to the bound derived by the LUX collaboration taken from Fig.~5 of Ref.~\cite{Akerib:2013tjd}. We also show our $90\%$ CL XENON100 bound. We assume the SHM with $| \bfv_\odot | = 245$ km/s (as in Ref.~\cite{Akerib:2013tjd}), and isospin-conserving SI couplings.}
\label{LUXcomparison}
\end{figure}

Fig.~\ref{LUXcomparison} compares the bound derived by the LUX collaboration (solid blue line, taken from Fig.~5 of Ref.~\cite{Akerib:2013tjd}) with the 90\% CL Maximum Gap limits we obtain assuming different observed events for WIMPs with SI isospin-conserving interactions (magenta lines, corresponding to, from the bottom, 0, 1, 3, 5 and 24 observed events). For this comparison, we use the same halo model as in Ref.~\cite{Akerib:2013tjd}, namely a SHM with $| \bfv_\odot | = 245$ km/s instead of our value $| \bfv_\odot | = 232$ km/s (see Section~\ref{sec:SHM}). We can see that the published LUX bound almost coincides with our 0 observed events limit for WIMP masses $m < 10$ GeV/$c^2$ and is between our 0 and 1 event limits for $m < 20$ GeV/$c^2$. The figure also shows that our 3, 5 and 24 event limits are conservative with respect to the LUX bound.

\section{The scattering rate for spin-independent interactions}

The WIMP-nucleus scattering rate within a detected energy interval $[\Ed_1, \Ed_2]$, expressed in counts/kg/day, is
\begin{multline}
\label{R}
R_{[\Ed_1, \Ed_2]}(t) =
\\
\frac{\rho}{\mDM} \sum_T \frac{C_T}{m_T} \int_0^\infty \ud \ER \, \int_{v \geqslant v_\text{min}(\ER)} \hspace{-18pt} \ud^3 v \, f(\bfv, t) \, v \, \frac{\ud \sigma_T}{\ud \ER}(\ER, \bfv)
\, \epsilon_2(\ER) \int_{\Ed_1}^{\Ed_2} \ud\Ed \, \epsilon_1(\Ed) G_T(\ER, \Ed).
\end{multline}
Here $\rho$ is the local DM density. $\mDM$ is the WIMP mass. The sum runs over all the different target nuclides $T$, of mass $m_T$, present in the detector with mass fraction $C_T$. $\vmin(\ER)$ is the minimum speed required for the incoming DM particle to cause a nuclear recoil with energy $\ER$,
\beq\label{vmin}
\vmin = \sqrt{\frac{m_T \ER}{2 \mu_T^2}}
\eeq
for an elastic collision, with $\mu_T = \mDM \, m_T / (\mDM + m_T)$ the WIMP-nucleus reduced mass. $v$ is the magnitude of the DM velocity vector $\bfv$. $f(\bfv, t)$ is the DM velocity distribution in Earth's frame, which is modulated in time due to Earth's rotation around the Sun; the time dependence of the rate \eqref{R} is then generally well approximated by the first terms of a harmonic series,
\beq\label{Rt}
R_{[\Ed_1, \Ed_2]}(t) = R^0_{[\Ed_1, \Ed_2]} + R^1_{[\Ed_1, \Ed_2]} \cos\!\left[ \omega (t - t_0) \right] ,
\eeq
where $t_0$ is the time of the maximum of the signal and $\omega = 2 \pi/$yr. $\ud \sigma_T / \ud \ER$ is the differential scattering cross section. $G_T(\ER, \Ed)$ is a resolution function, giving the probability that a recoil energy $\ER$ (usually quoted in keVnr for nuclear recoils) is measured as $\Ed$ (which can be expressed as an energy in keVee, for electron-equivalent, or as a number of photoelectrons, like S1); this function incorporates the mean value $\langle \Ed \rangle = Q_T(\ER) \ER$, with $Q_T$ the quenching factor, and the energy resolution $\sigma(\langle \Ed \rangle)$. Finally, $\epsilon_1(\Ed)$ and $\epsilon_2(\ER)$ are two counting efficiency or cut acceptance functions. For a derivation of Eq.~\eqref{R} and a slightly more thorough explanation of the quantities appearing in it, we refer the reader  \eg to Ref.~\cite{DelNobile:2013cva}.

The differential cross section for SI interactions is
\beq
\frac{\ud \sigma_T}{\ud \ER} = \sigma_T^{\rm SI}(\ER) \frac{m_T}{2 \mu_T^2 v^2} ,
\eeq
with
\beq\label{SIcrossection}
\sigma_T^{\rm SI}(\ER) = \sigma_p \frac{\mu_T^2}{\mu_p^2} [ Z_T + (A_T - Z_T)(f_n / f_p) ]^2 F_{{\rm SI}, T}^2(\ER) .
\eeq
Here $Z_T$ and $A_T$ are respectively the atomic and mass number of the target nuclide $T$, $F_{{\rm SI}, T}(\ER)$ is the nuclear SI form factor (which we take to be the Helm form factor~\cite{Helm:1956zz}), $f_n$ and $f_p$ are the effective DM couplings to neutrons and protons, $\sigma_p$ is the DM-proton cross section and $\mu_p$ is the DM-proton reduced mass. The isospin-conserving coupling  $f_n = f_p$ is  usually assumed by the experimental collaborations. The isospin-violating  coupling  $f_n / f_p = -0.7$~\cite{Kurylov:2003ra, Feng:2011vu} produces the maximum  cancellation in the expression inside the square bracket in Eq.~\eqref{SIcrossection} for  xenon, thus highly suppressing the interaction cross section. This suppression is phenomenologically interesting because it considerably weakens the bounds from xenon-based detectors such as XENON and LUX which provide some of the most restrictive bounds.

\section{The halo-dependent data comparison}
\label{sec:SHM}

A routine way to compare results from different direct DM detection experiments is by assuming a particular DM velocity distribution and DM density in Eq.~\eqref{R}. The velocity distribution in Earth's frame is related to the velocity distribution in the galactic frame $f_\text{G}({\bf u})$ by the Galilean transformation $f(\bfv, t) = f_\text{G}({\bf u} = \bfv + \bfv_\text{E}(t))$, where $\bfv_\text{E}(t)$ is Earth's velocity with respect to the galactic frame. The time dependence is due to the periodic motion of Earth around the Sun (see \eg Ref.~\cite{Savage:2008er} for details). The time average of $\bfv_\text{E}(t)$ over a period of one year corresponds to the Sun's velocity with respect to the galaxy $\bfv_\odot$.

In the SHM the DM velocity distribution in the galactic frame is  a truncated Maxwell-Boltzmann distribution,
\beq\label{f_G}
f_\text{G}({\bf u}) = \frac{\exp(- u^2 / v_0^2)}{(v_0 \sqrt{\pi})^3 N_\text{esc}} \, \theta(\vesc - u).
\eeq
Here $v_0$ is the most probable speed with respect to the galaxy and $\vesc$ is the local galactic escape speed beyond which the distribution is truncated via the step function $\theta(\vesc - u)$. The normalization factor $N_\text{esc} \equiv \text{erf}(v_\text{esc} / v_0) - 2 (v_\text{esc} / v_0) \exp(- v_\text{esc}^2 / v_0^2) / \sqrt{\pi}$ ensures the right normalization $\int \ud^3 u \, f_\text{G}({\bf u}) = 1$. When computing halo model dependent limits we use $| \bfv_\odot | = 232$ km/s~\cite{Savage:2008er}, $v_0 = 220$ km/s, $\vesc = 544$ km/s~\cite{Smith:2006ym}, and $\rho = 0.3$ GeV/$c^2$/cm$^3$ for the DM local density. 

Of particular importance for our considerations is the maximum WIMP speed $v_{\rm max}$ with respect to Earth, $v_{\rm max} = \vesc + | \bfv_\odot | $. There is an uncertainty  of less than 100  km/s  in  $| \bfv_\odot |$, stemming largely  from the uncertainty in the value of the  galactic rotation velocity at the position of the Sun or Local Standard of Rest  velocity (see \eg Ref.~\cite{Bozorgnia:2012eg} where extreme choices of this speed were shown to be  193 km/s and  324 km/s). The escape speed is, however, known with much smaller accuracy. The old usual value  of $\vesc$, many times still used in direct DM papers, is $v_{\rm esc}=650$~km/s.  However the estimates in the literature  vary,  starting with a minimum value of 400 km/s~\cite{Alexander-1982, Piffl:2013mla} to the until very recently most precise measurement by the Radial Velocity Experiment (RAVE)~\cite{Smith:2006ym} 2006 survey using high velocity stars,  544$^{+64}_{-46}$ km/s at the 90\% CL, with a median-likelihood value of 544 km/s which we use in this paper, to the most recent RAVE 2013 results~\cite{Piffl:2013mla}, which give a best estimate of 533$^{+54}_{-41}$ km/s at 90\% CL with an additional 4\% systematic uncertainty.  Recent theoretical papers, such as Refs.~\cite{Fornasa:2013iaa} and~\cite{Nesti:2013uwa}  mention larger upper values,  645 km/s  and 750 km/s at the 95\% CL respectively. A variation in $| \bfv_\odot |$ is expected to move the regions of interest and limits approximately in the same manner in parameter space, thus not changing significantly their relative position. However, a change in $\vesc$ can affect \ the limits on light WIMPs due to a heavy target like xenon more than other regions of interest and bounds obtained with lighter targets. Considering fixed $| \bfv_\odot |$ and taking extreme values among those just mentioned for  $\vesc$ alone, 400 km/s and 750 km/s, we get  $v_{\rm max} = 776^{+206}_{-144}$, \ie  a fractional change of about 30\% above and 20\% below the central value we assume for our SHM plots.

Once the DM velocity distribution and density are fixed, Eq.~\eqref{R} can be used to compute the rate in any detected energy interval $[\Ed_1, \Ed_2]$. The rate can be compared with the expected background and observed data to produce a bound on the parameter space of the model (in case the data are incompatible with a DM signal), or an allowed region (in case the data favor an interpretation in terms of a DM signal). For SI interactions one customarily chooses the DM-proton cross section $\sigma_p$ as the parameter to be constrained together with the mass $m$, as it does not depend on the detector and thus bounds and allowed regions from different experiments can be compared on the same plot.

We use the Maximum Gap Method~\cite{Yellin:2002xd} to produce $90\%$ CL bounds on the $m$--$\sigma_p$ parameter space from the CDMS-II-Ge, CDMS-II-Si, CDMSlite, XENON10, XENON100 and LUX experiments (we do not consider the CDMS-II-Ge modulation bound here). The SIMPLE bound is the $90\%$ CL Poisson limit and the CoGeNT 2011-2012 unmodulated rate bound is the 90\% CL limit (in a raster scan). We compute the $68\%$ and $90\%$ CL allowed regions from the  DAMA (sodium only) and CoGeNT 2011-2012 modulation data~\cite{CoGeNT2011release, Aalseth:2012if} and from the CDMS-II-Si unmodulated signal using the Extended Maximum Likelihood method~\cite{Barlow:1990vc}. We also include the low mass CRESST-II $2 \sigma$ CL allowed region in the $m$--$\sigma_p$ parameter space from Fig.~14 of the published version of~\cite{Angloher:2011uu}. We take the $90\%$ CL maximum likelihood CoGeNT 2014 regions (with and without a floating surface background energy distribution) and upper limit (with unconstrained neutron background) from Fig.~11 of Ref.~\cite{Aalseth:2014jpa}. We can directly compare the regions and bound taken from the papers of CRESST-II and CoGeNT with the others in our plots because the SHM parameters they adopted are very similar to ours. We disregard the iodine contribution in DAMA and the tungsten contribution in CRESST-II, as they are both negligible at the low DM masses we are interested in. The low mass CRESST-II region comes from WIMP scattering off O and Ca. Since more than 99\% of the O and Ca isotopes have an equal number of protons and neutrons, the O and Ca SI cross sections scale as $(f_n + f_p)^2$.  Hence we can translate the published CRESST-II region for  isospin-conserving interactions to the region for isospin-violating interactions  by dividing the allowed $\sigma_p$  range for each mass by a factor of $[(f_n + f_p) / (2 f_p)]^2 = (1 + f_n / f_p)^2 / 4$, which is 1/44.4 for $f_n/f_p=-0.7$.  The corresponding factor to translate the CoGeNT 2014 regions and bound is $\left\{\sum_T[1 + (f_n/f_p) (A_T-Z_T)/Z_T]^2 (C_T/m_T)\right\}/\left\{\sum_T[1+ (A_T-Z_T)/Z_T]^2 (C_T/m_T) \right\}$ where the summations are over all Ge isotopes, which amounts to $1/384.8$ for $f_n/f_p=-0.7$.

\begin{figure}[t]
\centering
\includegraphics[width=0.49\textwidth]{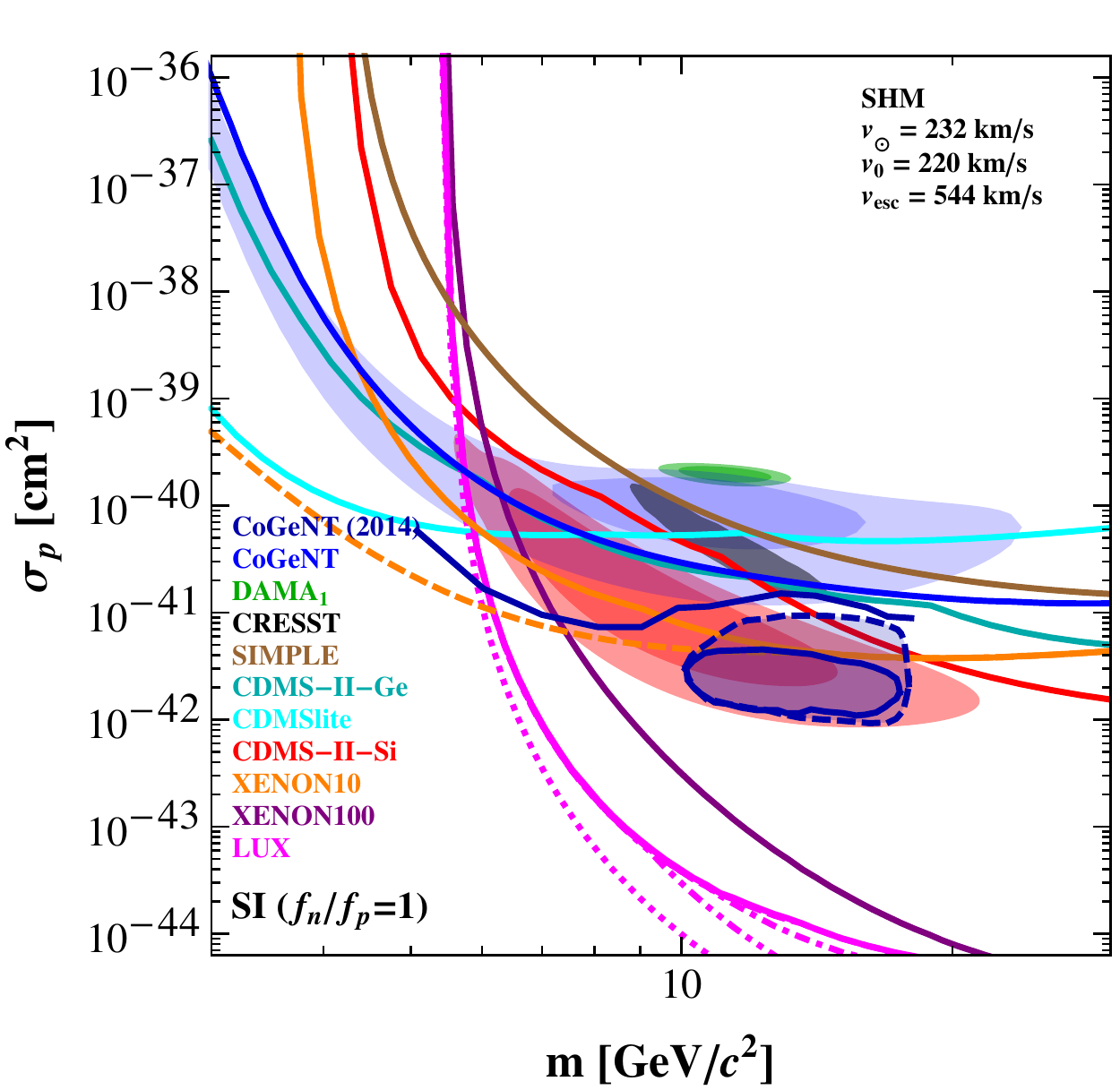}
\includegraphics[width=0.49\textwidth]{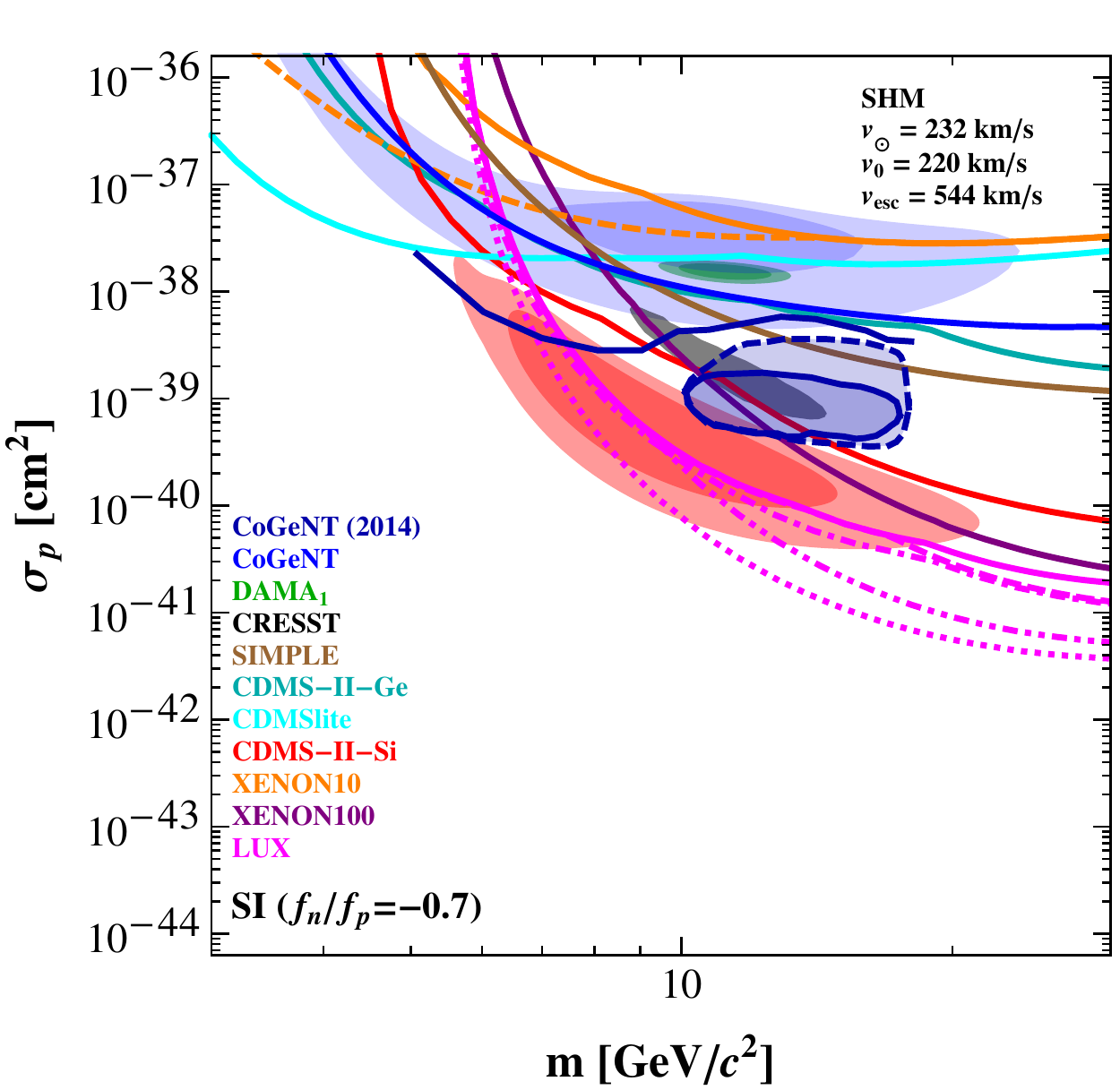}
\caption{$90\%$ CL bounds and $68\%$ and $90\%$ CL allowed regions in the SI DM-proton cross section vs WIMP mass plane, assuming the SHM. The CRESST-II low mass allowed region is only shown at $2 \sigma$ CL. The CoGeNT 2014 solid and dashed dark blue contours show $90\%$ regions, with fixed and floating surface event background energy distributions, respectively. The left panel is for isospin-conserving $f_n = f_p$ couplings, the right panel is for isospin-violating $f_n / f_p = -0.7$ couplings. For XENON10 (orange bounds), the solid line is produced by conservatively setting the electron yield $\mathcal{Q}_{\rm y}$ to zero below 1.4 keVnr as in Ref.~\cite{Aprile:2012nq}, while the dashed line ignores the $\mathcal{Q}_{\rm y}$ cut. For LUX (magenta bounds), the limits correspond to (from the bottom) 0, 1, 3, 5, and 24 observed events (see Section~\ref{data} and Fig.~\ref{LUXcomparison}). Only sodium is considered for DAMA, and the quenching factor is taken to be $Q_{\rm Na} = 0.3$.}
\label{m-sigma}
\end{figure}

Fig.~\ref{m-sigma} shows our results for isospin-conserving interactions (left panel) and isospin-violating interactions (right panel). We include the CoGeNT 2011-2012 (labeled simply ``CoGeNT" in the figures) and DAMA modulation regions (the latter for scattering off Na only), the CoGeNT 2014 regions (derived in Ref.~\cite{Aalseth:2014jpa} from both the average rate and the modulation amplitude), the CRESST-II (taken from Ref.~\cite{Angloher:2011uu}) and CDMS-II-Si DM-signal regions, and the SIMPLE, CoGeNT 2011-2012 average rate, CoGeNT 2014, CDMS-II-Ge, CDMS-II-Si, CDMSlite, XENON10, XENON100, and LUX upper bounds. For XENON10, we show two bounds: one derived setting and the other without setting the electron yield $\mathcal{Q}_{\rm y}$ to zero below 1.4 keVnr. For LUX we show the bounds corresponding to 0, 1, 3, 5, and 24 observed events, as in Fig.~\ref{LUXcomparison}.
For isospin-conserving SI interactions, the LUX and CDMSlite upper bounds together exclude the DAMA, CoGeNT, CRESST-II, and CDMS-II-Si DM-signal regions (in this SHM analysis). LUX provides the strongest limit for WIMP masses above 5.5 GeV/$c^2$. However, the imposed conservative threshold of 3.0 keVnr for the scintillation signal, combined with our choice of escape velocity from the galaxy, does not allow the LUX analysis (nor the XENON100 analysis) to be sensitive below 5.5 GeV/$c^2$. The CDMSlite limit, instead, extends to lower WIMP masses: it is less stringent than the XENON10 limit without the conservative threshold of 1.4 keVnr on the ionization signal, becoming comparable to it at masses below 4 GeV/$c^2$; it is stronger than the XENON10 limit with 1.4 keVnr threshold for WIMP masses below 5.5 GeV/$c^2$.

For isospin-violating SI interactions with $f_n/f_p=-0.7$, the combination of LUX and CDMSlite upper limits exclude the DAMA, CoGeNT, and CRESST-II (but not CDMS-II-Si) DM-signal regions in this SHM analysis. CDMSlite provides the strongest bound at WIMP masses below $\sim 6$ GeV/$c^2$. The LUX limit, despite the isospin-violating arrangement to suppress interactions with xenon, strongly excludes the DAMA and CRESST-II regions, and portions of the CoGeNT and CDMS-II-Si regions. 

Notice that the CoGeNT 2011-2012 modulation region is largely incompatible with the upper limit imposed by the average rate of the same data, when assuming the SHM for the dark halo of our galaxy. The  new CoGeNT 2014 region from Ref.~\cite{Aalseth:2014jpa} is below both the limit based on the CoGeNT 2011-2012 rate (that we derived) and the CoGeNT 2014 limit from Ref.~\cite{Aalseth:2014jpa}. This CoGeNT 2014 region partially overlaps with the CDMS-Si region, and for isospin-violating SI interactions also with the CRESST-II region,  but is rejected by the LUX limit.

The lowest reach of the LUX and XENON100 limits, call it $m_{\rm limit}$,  is determined by the smallest measurable recoil energy, here 3 keVnr for both, \ie by the minimum WIMP mass for which the maximum possible xenon recoil energy for a WIMP moving at the maximum possible speed with respect to Earth, $v_{\rm max}$, is  3 keVnr. Since this energy depends on $\mu_T^2 v_{\rm max}^2 \simeq m_{\rm limit}^2  v_{\rm max}^2$ for $m_{\rm limit} \ll m_T$ we have $m_{\rm limit} \sim v_{\rm max}^{-1}$. Given the uncertainty in $v_{\rm max}$ of about 30\% above and 20\% below the value we assume for our SHM plots, as argued after Eq.~\eqref{f_G},  the position of the lowest reach of the LUX and XENON100 limits can change from about 4.6 GeV/$c^2$ to 7.5 GeV/$c^2$.  The signal regions do not depend so strongly on the escape speed value. Thus, because of the CDMSlite limit the conclusions of our SHM analysis do not change due to the uncertainties in $v_{\rm max}$.

This SHM analysis of SI interactions puts the DM interpretation of the DAMA annual modulations in severe tension with the LUX and CDMSlite results, and the upper limit of CoGeNT. In the isospin-violating case we consider, the CoGeNT 2014 region is incompatible only with the LUX limit, although the XENON100 and CDMS-II-Si limits constrain it significantly too. The analysis in Ref.~\cite{Gresham:2013mua} (which does not include the CoGeNT 2014 data), although less conservative than ours because it is based on taking a signal fraction of 1/2 below the NR mean line (see our discussion in Section~\ref{data}), reaches qualitatively similar conclusions.

\section{The halo-independent data comparison}
Given the large uncertainties on the properties of the DM halo, most notably on the velocity distribution $f_\text{G}(\bfv)$, a method was put forth in Ref.~\cite{Fox:2010bz, Frandsen:2011gi, Gondolo:2012rs} to compare the different experimental results in a halo-independent manner. By changing integration variable from $\ER$ to $\vmin$ through \Eq{vmin}, we can rewrite \Eq{R} as
\begin{align}
\label{R1}
R^{\rm SI}_{[\Ed_1, \Ed_2]}(t) & = \int_0^\infty \ud \vmin \, \tilde{\eta}(\vmin, t) \, \eR^{\rm SI}_{[\Ed_1, \Ed_2]}(\vmin) ,
\end{align}
where the velocity integral $\tilde{\eta}$ is 
\beq
\label{eta0}
\tilde{\eta}(\vmin, t) \equiv \frac{\rho \sigma_p}{\mDM} \int_{v \geqslant \vmin} \ud^3 v \, \frac{f(\bfv, t)}{v} ,
\eeq
and the response function $\eR^{\rm SI}_{[\Ed_1, \Ed_2]}(\vmin)$ for WIMPs with SI interactions is defined as
\beq
\eR^{\rm SI}_{[\Ed_1, \Ed_2]}(\vmin) \equiv
2 \vmin \sum_T \frac{C_T}{m_T} \frac{\sigma_T^{\rm SI}(\ER(\vmin))}{\sigma_p}
\, \epsilon_2(\ER) \int_{\Ed_1}^{\Ed_2} \ud\Ed \, \epsilon_1(\Ed) G_T(\ER(\vmin), \Ed).
\eeq

Due to the revolution of Earth around the Sun, the velocity integral $\tilde{\eta}(\vmin,t)$ has an annual modulation in the same way as the rate \eqref{Rt}:
\beq\label{etat}
\tilde{\eta}(\vmin, t) \simeq \tilde{\eta}^0(\vmin) + \tilde{\eta}^1(\vmin) \cos\!\left[ \omega (t - t_0) \right] .
\eeq
The unmodulated and modulated components $\tilde{\eta}^0$ and $\tilde{\eta}^1$ enter respectively in the definition of unmodulated and modulated parts of the rate, $R^0_{[\Ed_1, \Ed_2]}$ and $R^1_{[\Ed_1, \Ed_2]}$. The functions $\tilde{\eta}^0(\vmin)$ and $\tilde{\eta}^1(\vmin)$, once the WIMP mass and interactions are fixed, are detector-independent quantities. We can therefore map the rate measurements and bounds of different experiments into measurements of and bounds on $\tilde{\eta}^0(\vmin)$ and $\tilde{\eta}^1(\vmin)$, as functions of $\vmin$.

For experiments with putative DM signals, in light of Eq.~\eqref{R1} we may interpret the measured rates $\hat{R}^{\, i}_{[\Ed_1, \Ed_2]} \pm \Delta{R}^{\, i}_{[\Ed_1, \Ed_2]}$ in an energy interval $[\Ed_1, \Ed_2]$ as averages of the $\tilde{\eta}^i(\vmin)$ functions weighted by the response function $\eR^{\rm SI}_{[\Ed_1, \Ed_2]}(\vmin)$:
\beq
\label{avereta}
\overline{\tilde{\eta}^{\, i}_{[\Ed_1, \Ed_2]}} \equiv \frac{\hat{R}^{\, i}_{[\Ed_1, \Ed_2]}}
{\int \ud\vmin \, \eR^{\rm SI}_{[\Ed_1, \Ed_2]}(\vmin)} .
\eeq
Each such average corresponds to a point with error bars in the $(\vmin, \tilde{\eta})$ plane. The vertical bars  are given by $\Delta\overline{\tilde{\eta}^{\, i}_{[\Ed_1, \Ed_2]}}$ computed by replacing $\hat{R}^{\, i}_{[\Ed_1, \Ed_2]}$ by  $\Delta{R}^{\, i}_{[\Ed_1, \Ed_2]}$  in Eq.~\eqref{avereta}.  The  $\Delta{R}^{\, i}$ we use correspond to  the $68\%$ confidence interval. The horizontal bar shows the $\vmin$ interval where the response function $\eR^{\rm SI}_{[\Ed_1, \Ed_2]}(\vmin)$ for the given experiment is sufficiently different from zero. Following Refs.~\cite{Gondolo:2012rs}  and~\cite{DelNobile:2013cta} we choose  the  horizontal  bar  extending over the interval $[{\vmin}_{,1}, {\vmin}_{,2}] = [v_{\rm min}(\Ed_1 - \sigma(\Ed_1)), v_{\rm min}(\Ed_2 + \sigma(\Ed_2))]$, where $\sigma(\Ed)$ is the energy resolution and the function $v_{\rm min}(\Ed)$ is obtained from $v_{\rm min}(\ER)$ in Eq.~\eqref{vmin} by using the recoil energy  $\ER$ that produces the mean $\langle \Ed \rangle$ equal to  the measured energy $\Ed$. When isotopes of the same element are present, like for Xe or Ge, the $v_{\rm min}$ intervals of the different isotopes almost completely overlap, and we take $v_{\rm min,1}$ and $v_{\rm min,2}$ to be the $C_T$-weighted averages over the isotopes of the element. When there are nuclides belonging to very different elements, like Ca and O in CRESST-II, a more complicated procedure should be followed (see Ref.~\cite{Gondolo:2012rs} for details). We use this procedure for the modulation data of DAMA and CoGeNT, and  the rate measurements of CoGeNT (without subtracting a flat background), CRESST-II and CDMS-II-Si. 
We also use an almost identical procedure to compute the upper limit over the interval $[{\vmin}_{,1}, {\vmin}_{,2}]$ for the negative result of the CDMS-II-Ge modulation analysis, which specifically constrains $\tilde{\eta}^1$.

\begin{figure}[!htbp]
\centering
\includegraphics[width=0.46\textwidth]{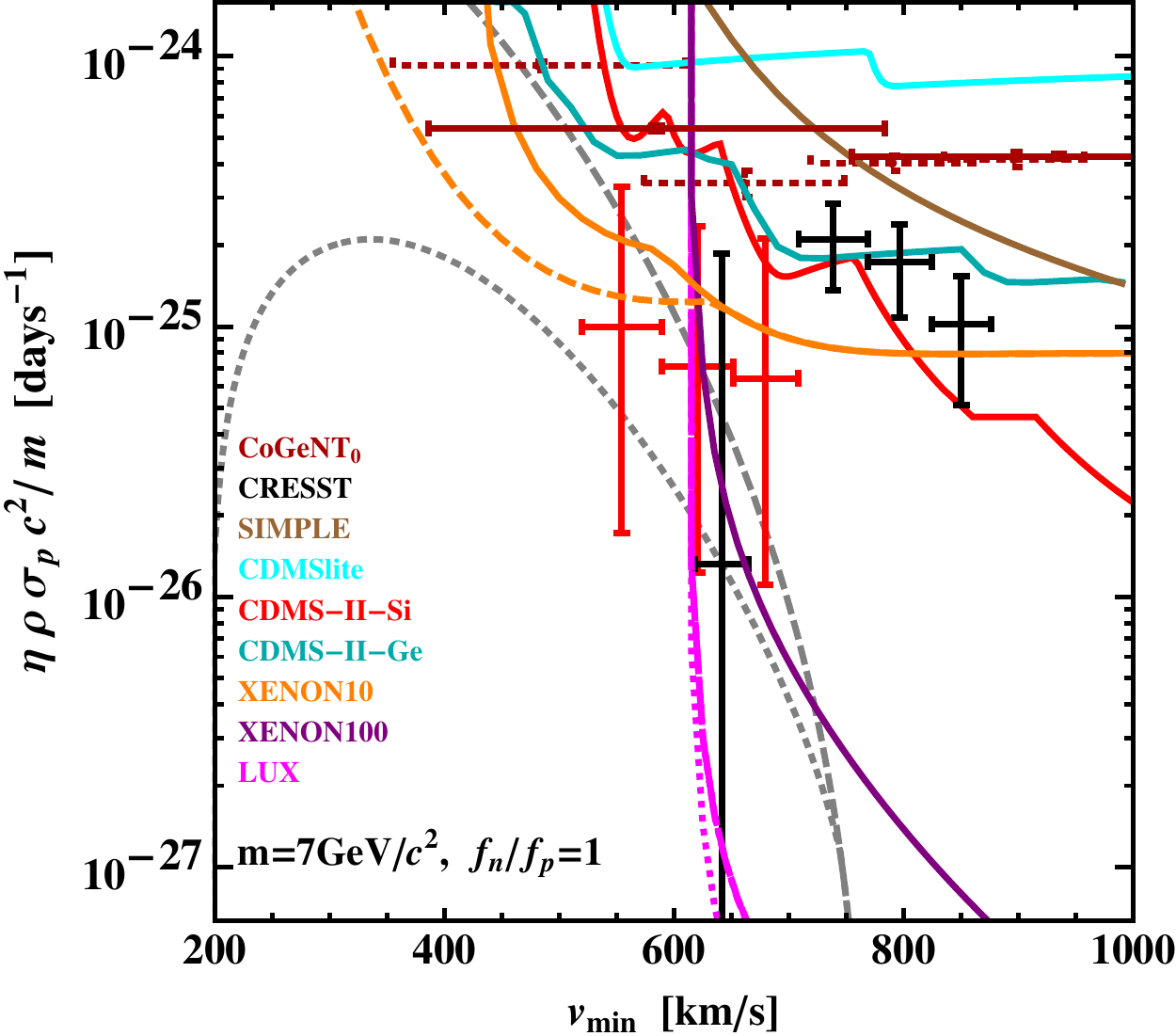}
\includegraphics[width=0.46\textwidth]{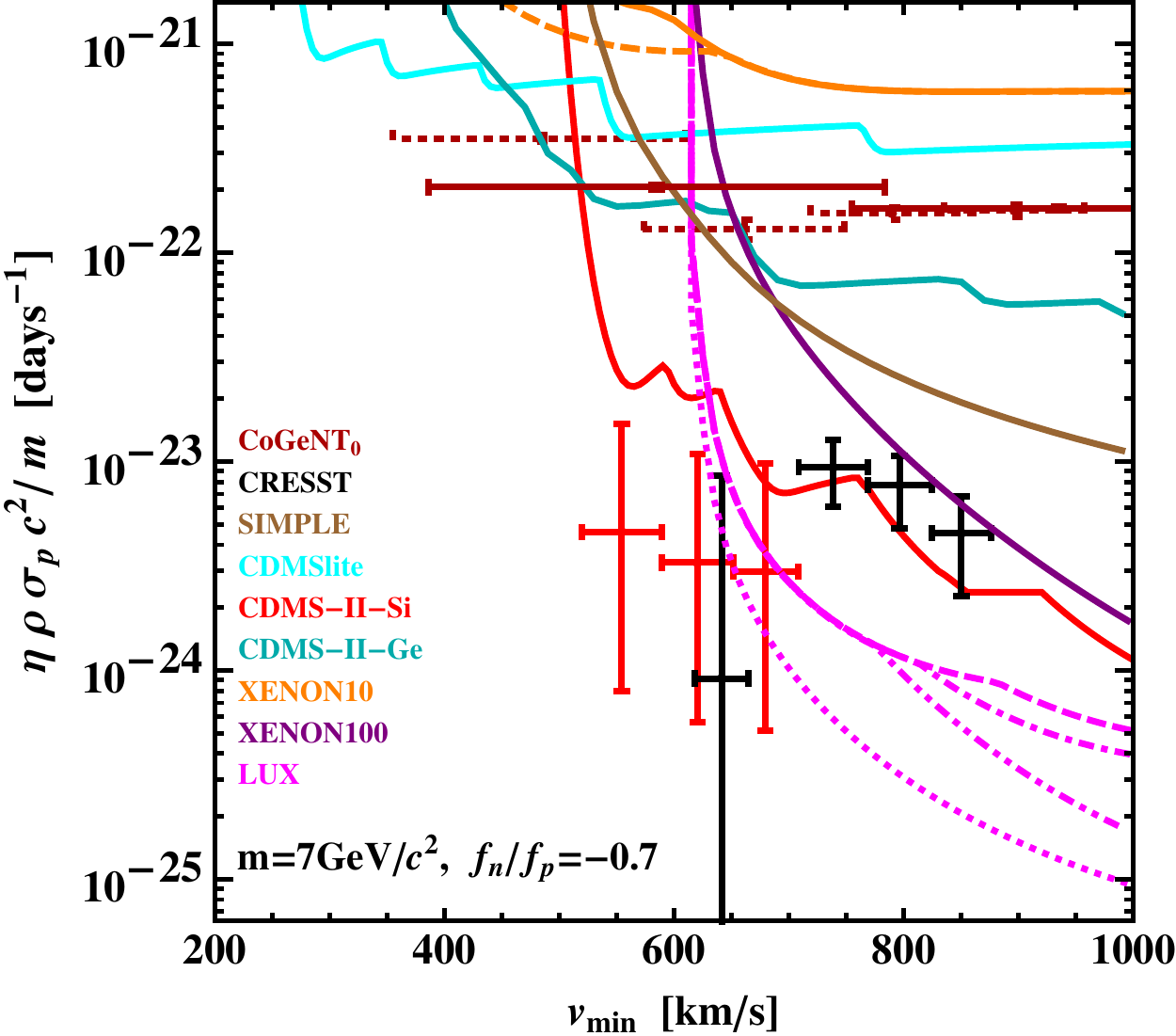}
\\
\includegraphics[width=0.46\textwidth]{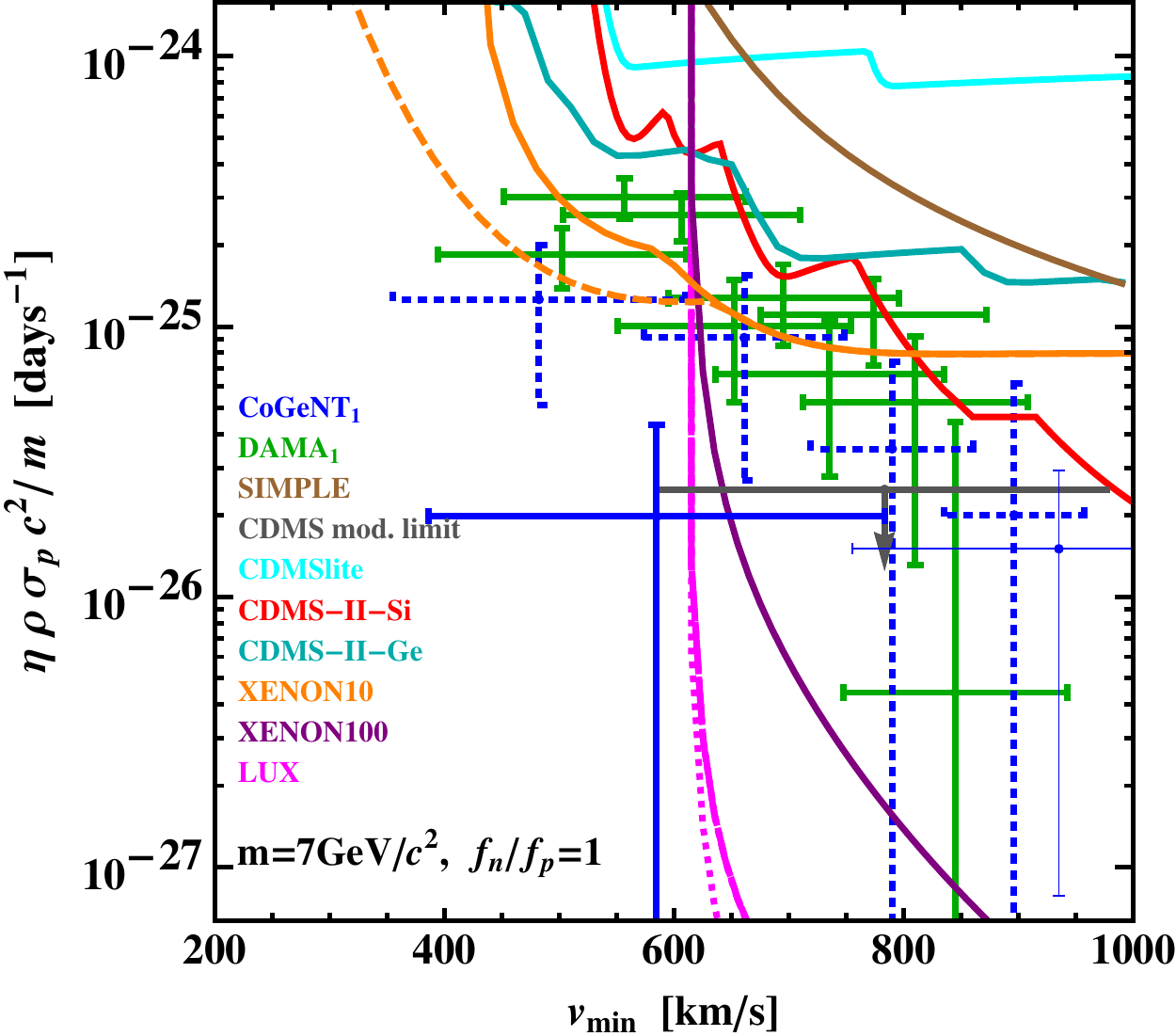}
\includegraphics[width=0.46\textwidth]{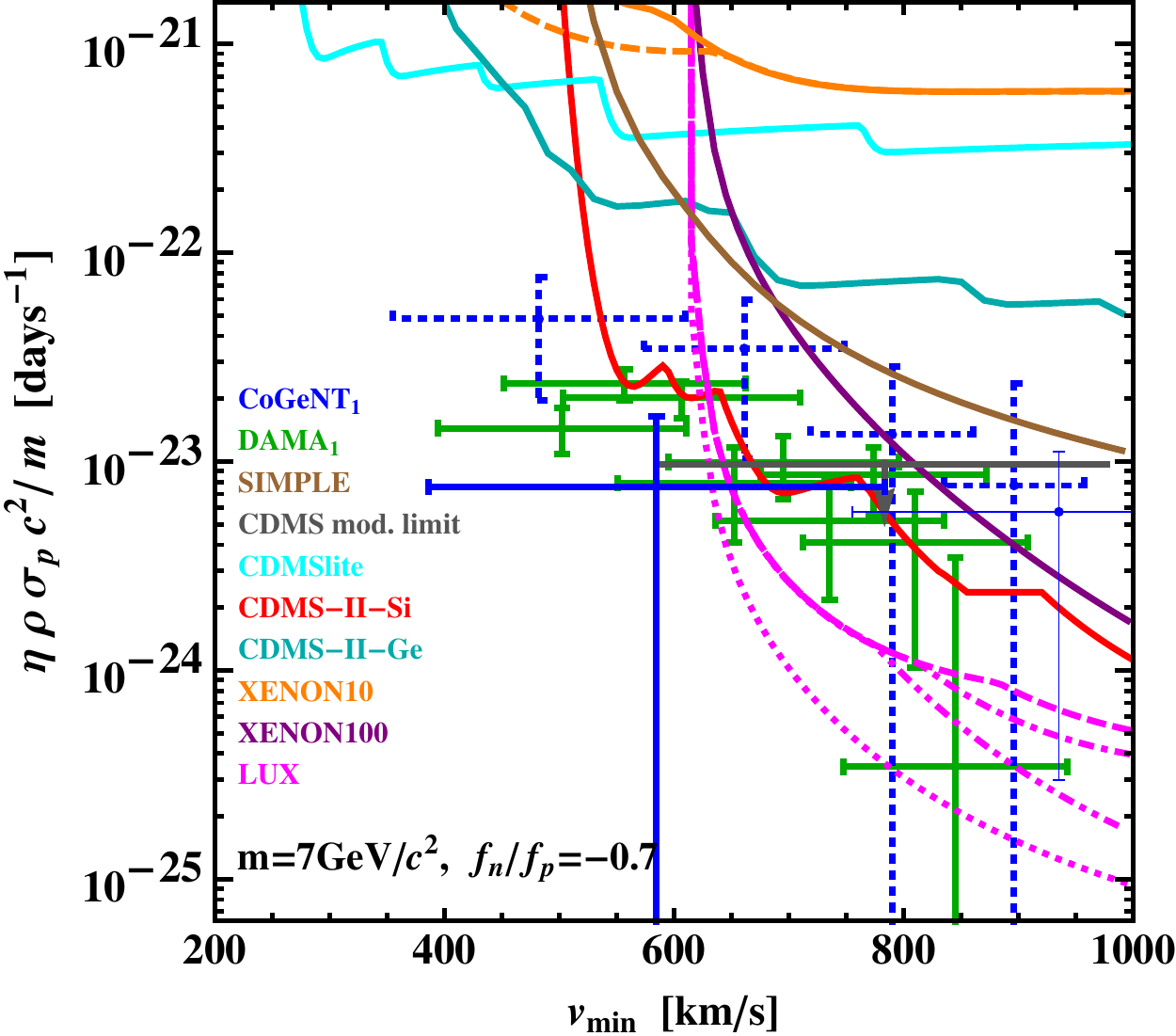}
\\
\includegraphics[width=0.46\textwidth]{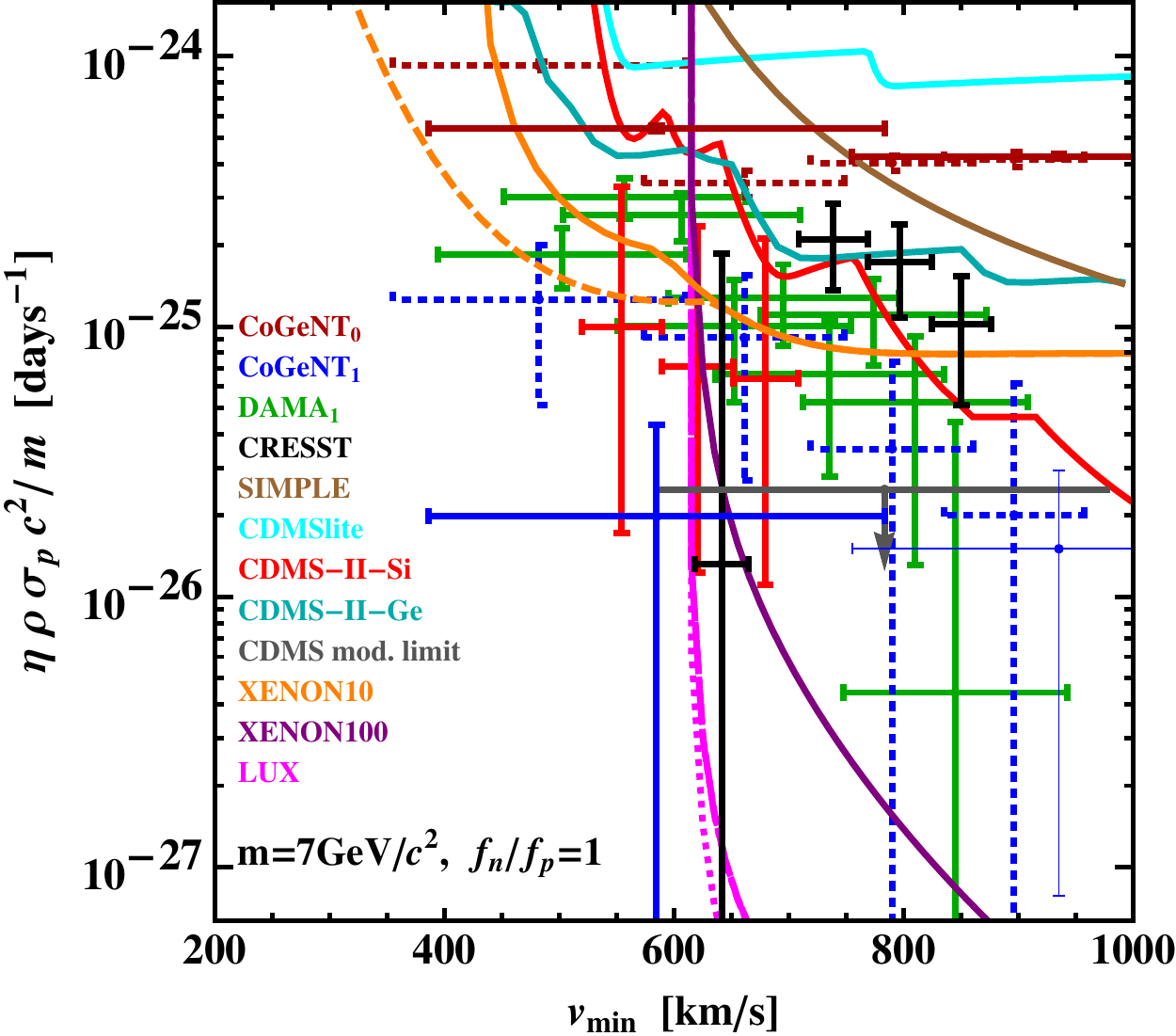}
\includegraphics[width=0.46\textwidth]{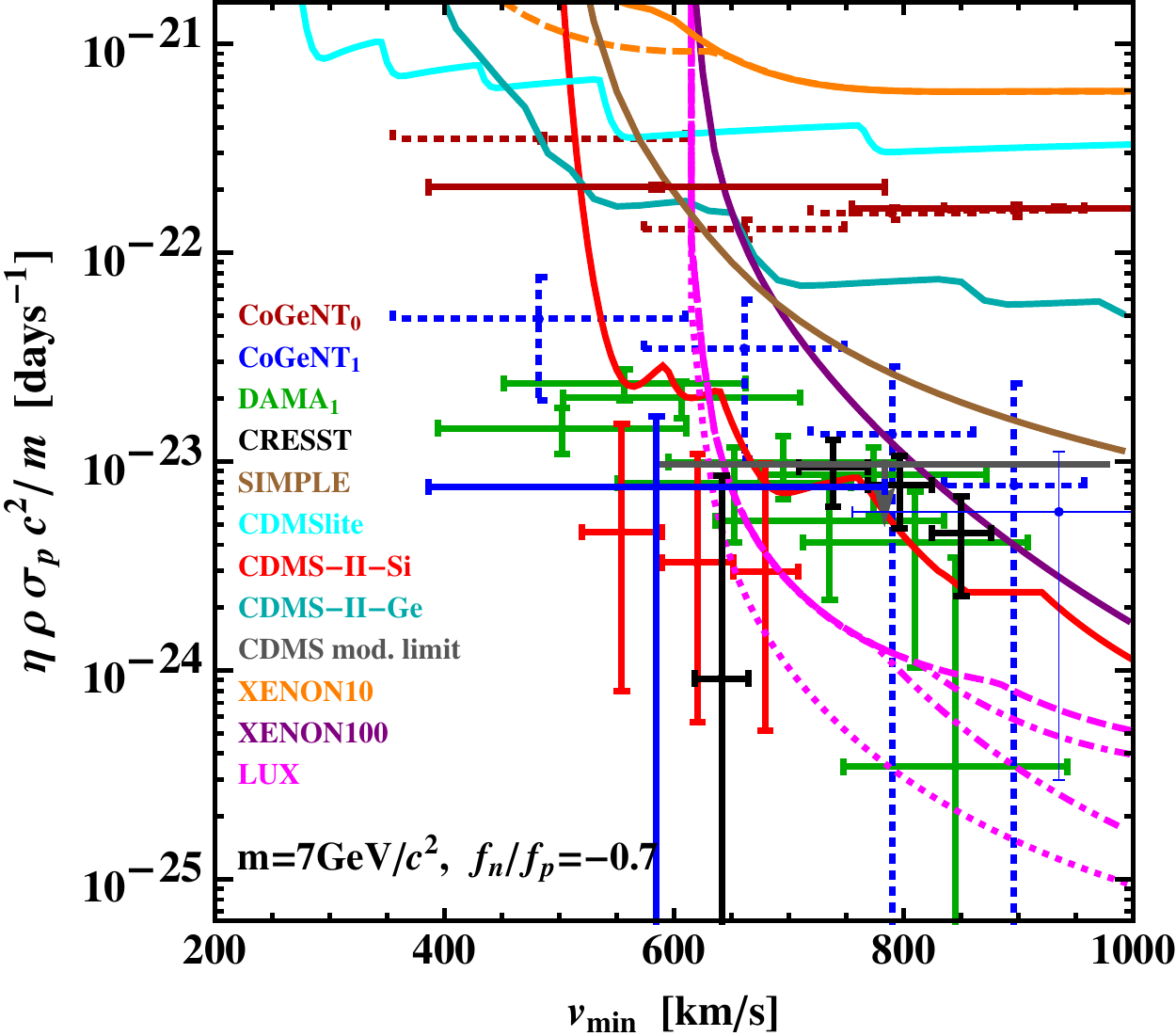}
\caption{Measurements of and bounds on $\tilde\eta c^2$ for $m = 7$ GeV/$c^2$. The left and right columns are for isospin-conserving  and isospin-violating interactions, respectively. The top, middle and bottom rows show measurements and bounds for the unmodulated component $\tilde\eta^0 c^2$, for the modulated component $\tilde\eta^1 c^2$, and for both together, respectively. The middle row also shows the upper bounds on $\tilde\eta^0 c^2$ from the plots on the top row. The dashed gray lines in the top left panel show the SHM $\tilde{\eta}^0 c^2$ (upper line) and $\tilde{\eta}^1 c^2$ (lower line) for $\sigma_p = 10^{-40}$ cm$^2$, which provides a good fit to the CDMS-II-Si data.
}
\label{eta7}
\end{figure}

\begin{figure}[!htbp]
\centering
\includegraphics[width=0.46\textwidth]{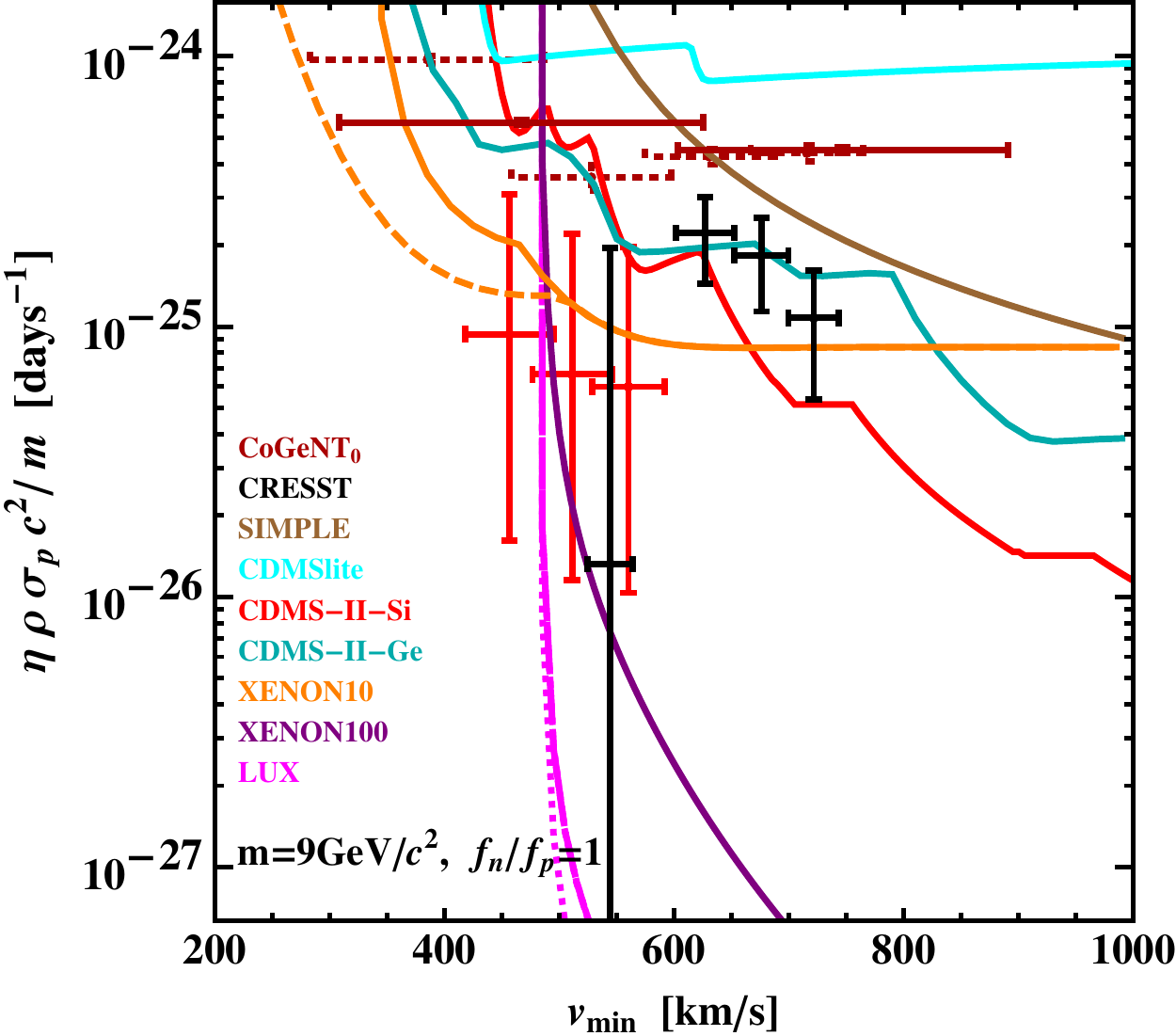}
\includegraphics[width=0.46\textwidth]{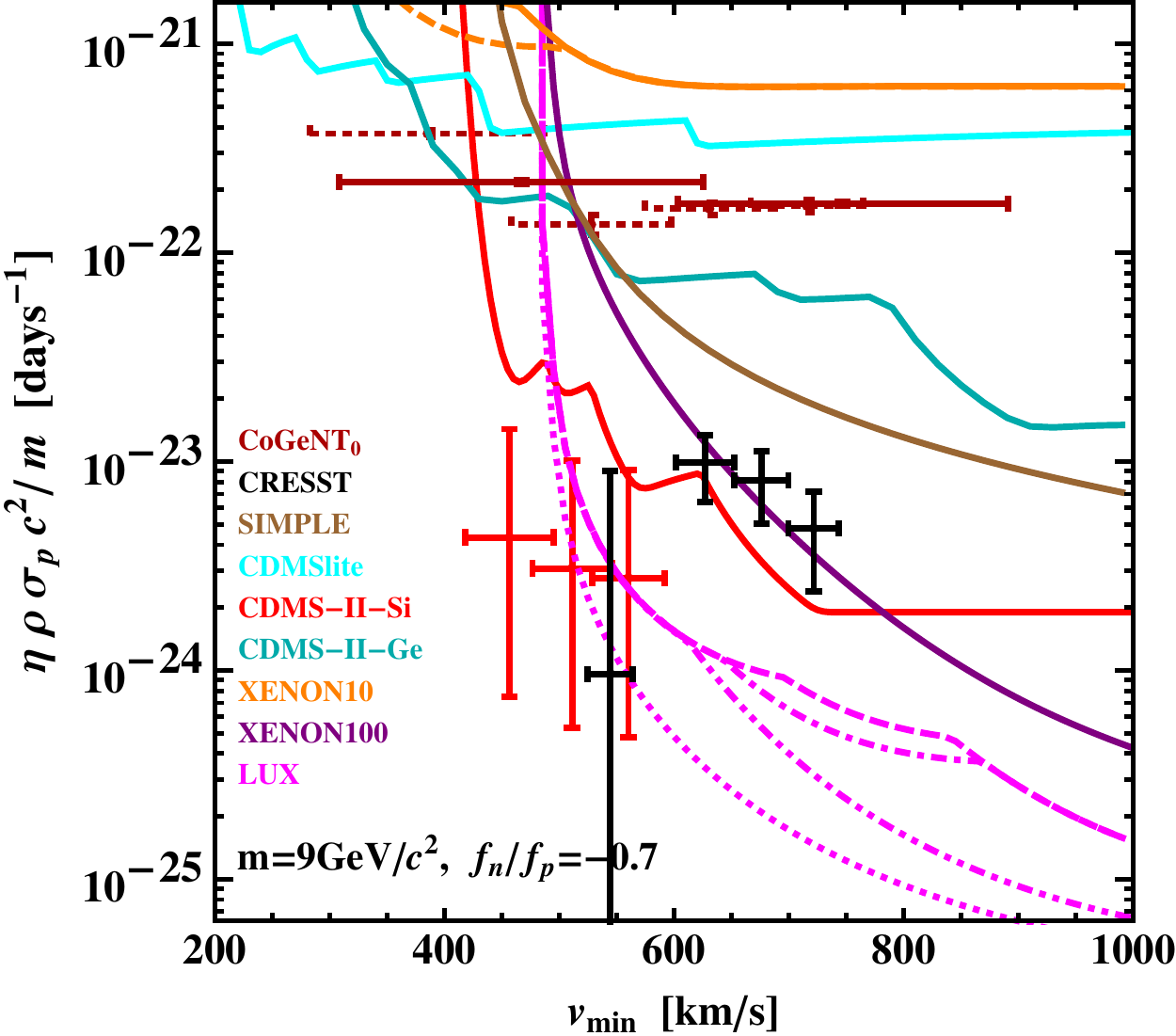}
\\
\includegraphics[width=0.46\textwidth]{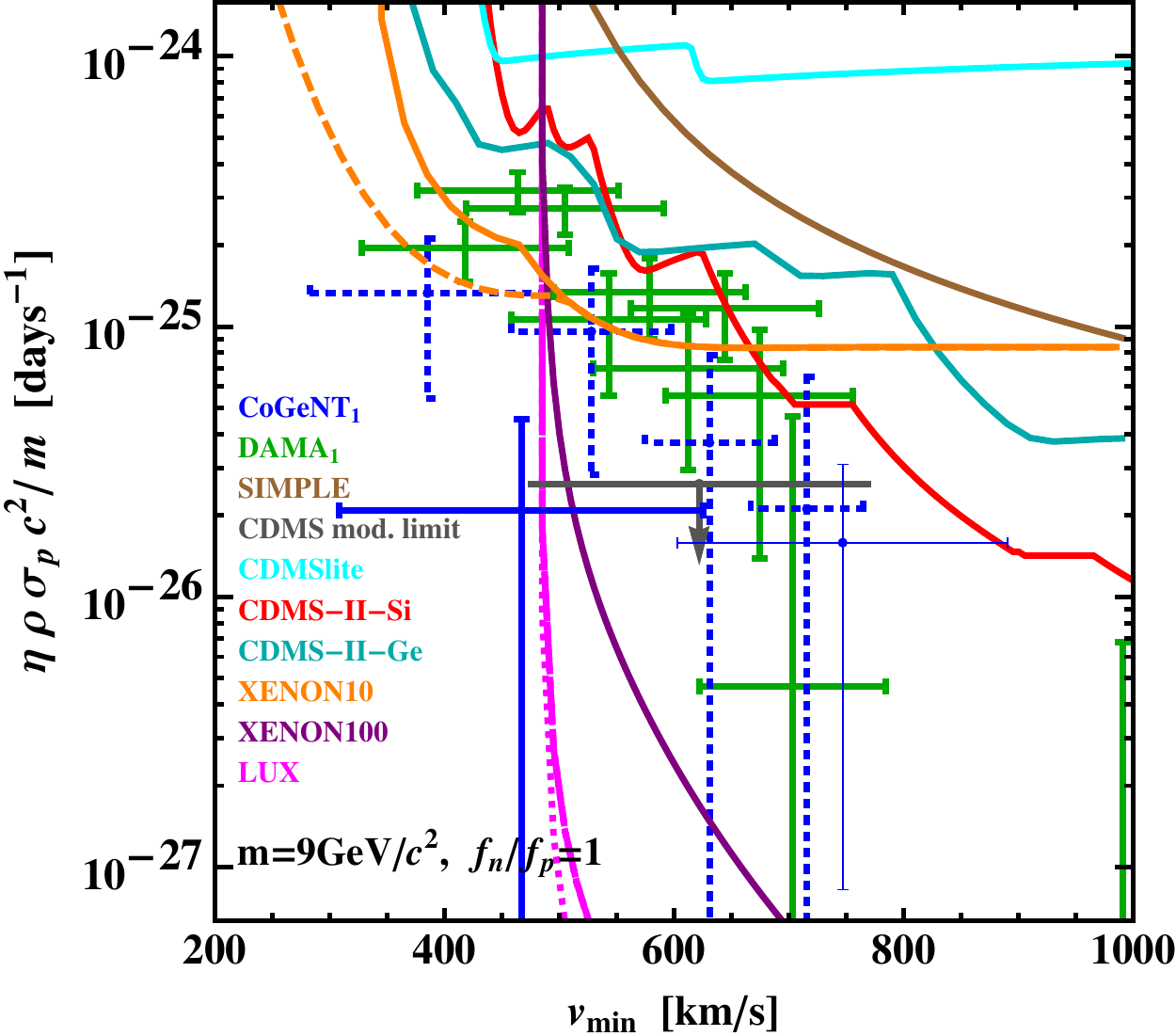}
\includegraphics[width=0.46\textwidth]{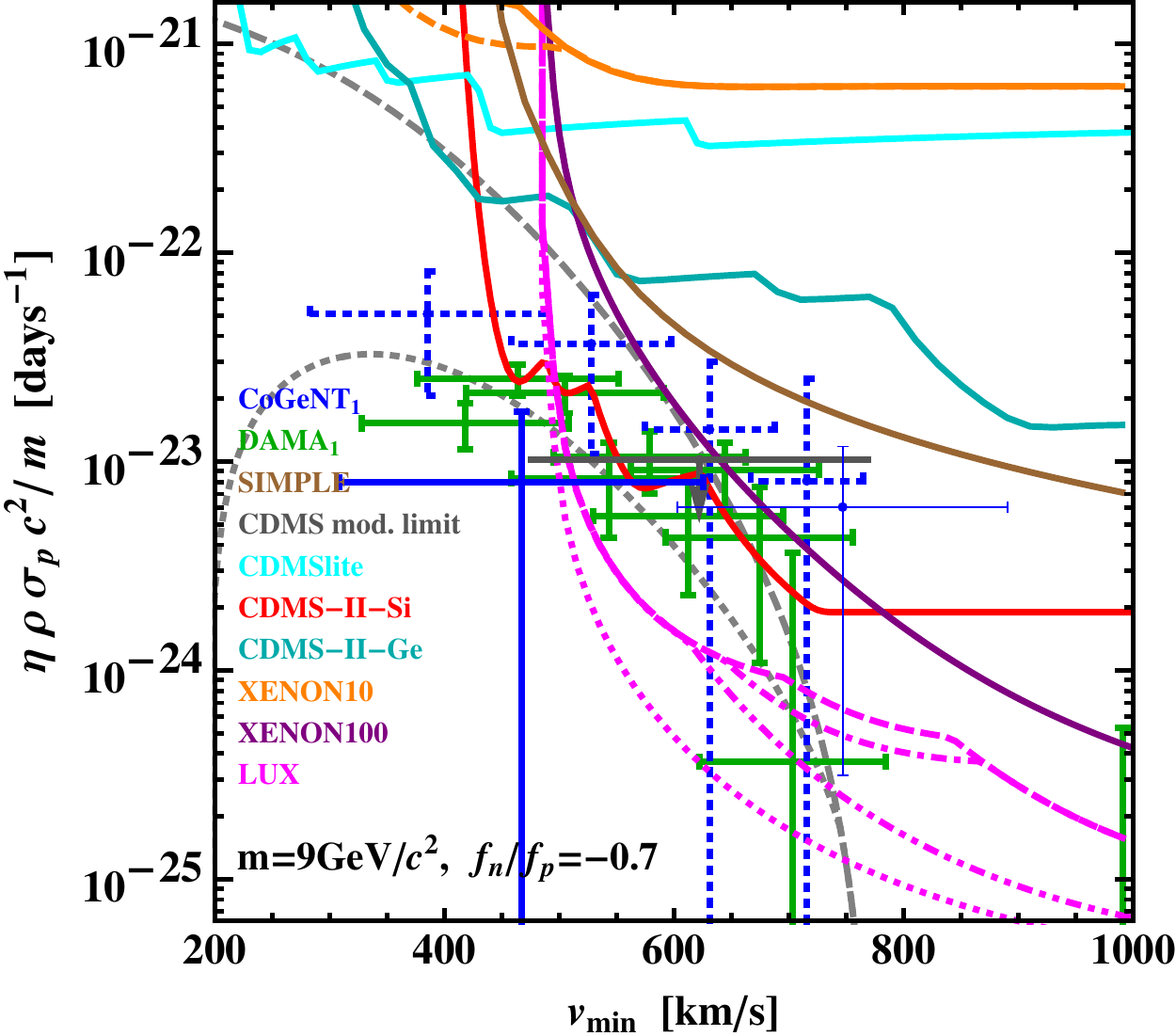}
\\
\includegraphics[width=0.46\textwidth]{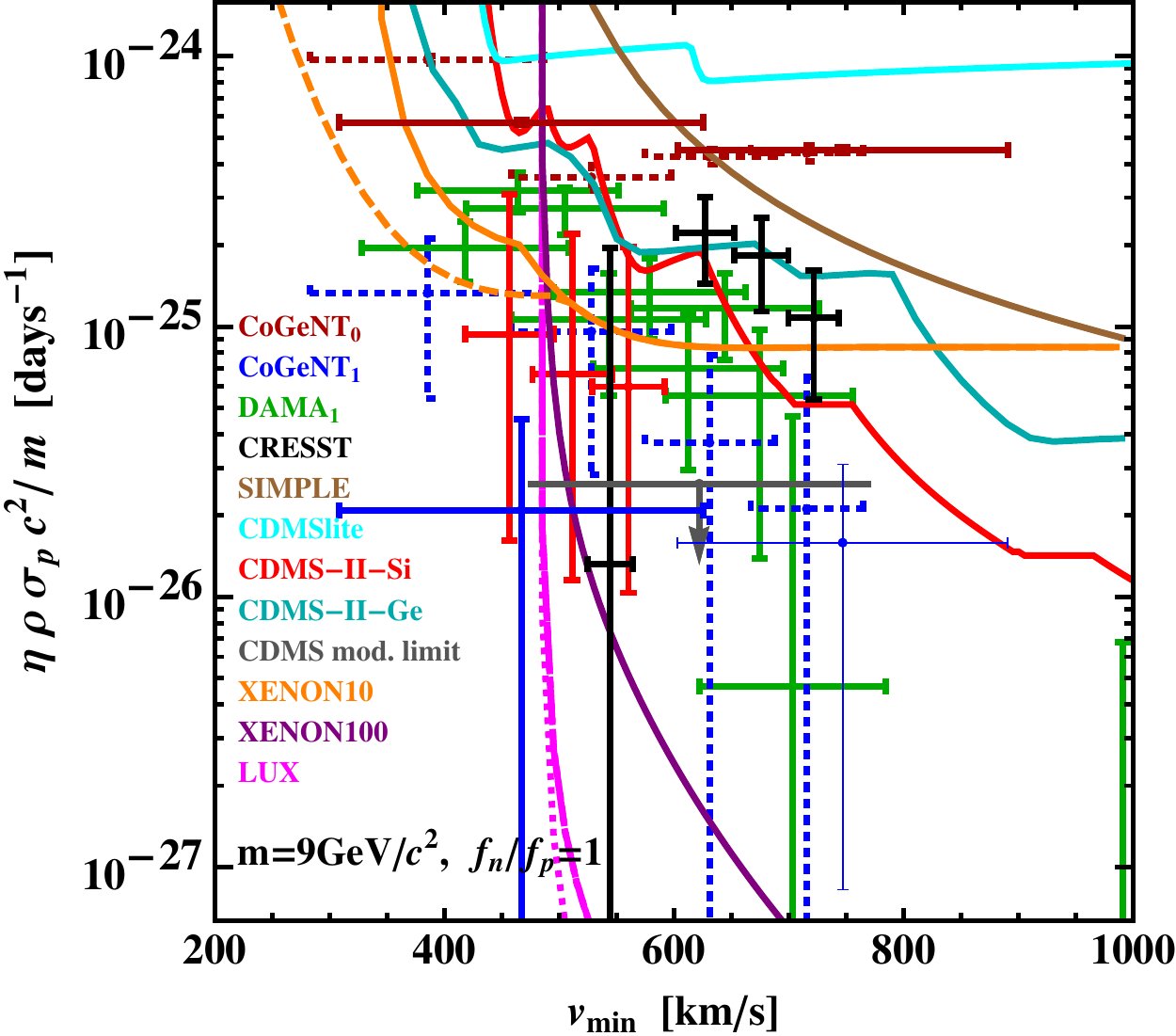}
\includegraphics[width=0.46\textwidth]{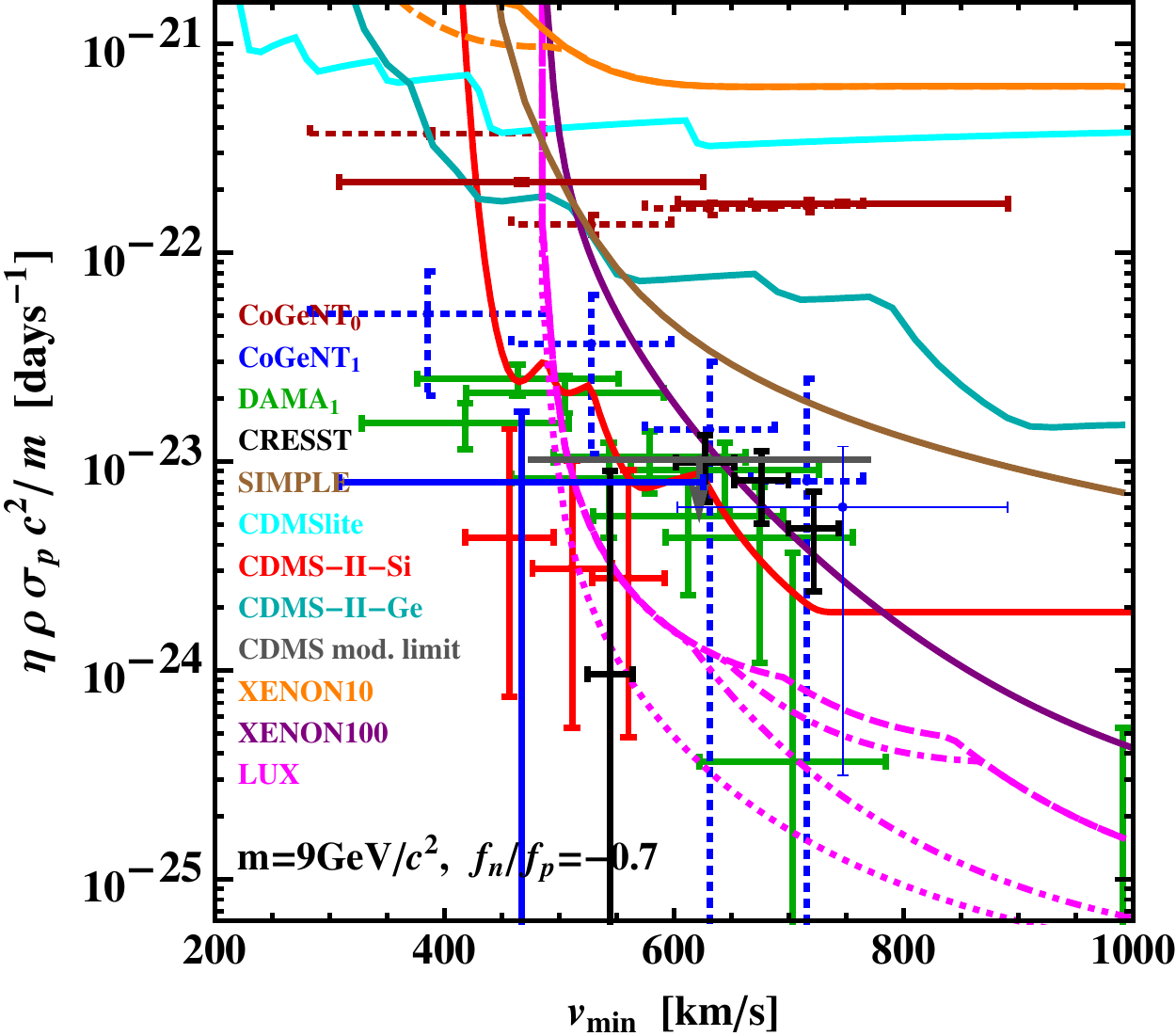}
\caption{Same as Fig.~\ref{eta7}, but for a DM mass $m = 9$ GeV/$c^2$. The dashed gray lines in the middle right panel show the expected $\tilde{\eta}^0 c^2$ (upper line) and $\tilde{\eta}^1 c^2$ (lower line) for a WIMP-proton cross section $\sigma_p = 2 \times 10^{-38}$ cm$^2$ in the SHM, which gives a good fit to the DAMA modulation data.}
\label{eta9}
\end{figure}

We analyze the CoGeNT 2014 data~\cite{Aalseth:2014eft} in the same way we analyze the older data~\cite{Gondolo:2012rs,DelNobile:2013cta}, following the procedure  specified in Ref.~\cite{Aalseth:2012if}, except that instead of using four energy bins (0.425 to 1.1125 keVee, 1.1125 to 1.8 keVee, 1.8 to 2.4875 keVee, and 2.4875 to 3.175 keVee) we use just two (0.5 to 2.0 keVee and 2.0 keVee up to 4.5 keVee), the same used in Ref.~\cite{Aalseth:2014eft}. For the residual surface event correction, $C(E)=1-\exp(-aE)$ with $E$ in keVee, we used $a=1.21$~\cite{Aalseth:2012if}. The  crosses corresponding to the older CoGeNT data are indicated with dashed lines in Figs.~\ref{eta7} and \ref{eta9}, and those of the CoGeNT 2014 data with solid lines. With our choice of modulation phase (DAMA's best fit value of 152.4 days from January 1st) and modulation period of one year in our fit to the data~\cite{Gondolo:2012rs}, the modulation amplitude we find in the  2.0 -- 4.5 keVee bin is negative, thus we show its magnitude in the figures with a thinner solid line. However, in both bins the modulation amplitude is compatible with zero, at the $0.9\sigma$ and $1.1\sigma$ level for the first and second, respectively. We checked that in each bin the reduced chi-square of the best fit values of the average rate and the modulation amplitude is close to one.  Since the vertical bars of the crosses in Figs.~\ref{eta7} and~\ref{eta9} represent $1\sigma$ interval, the first cross extends to zero.

Upper bounds on the unmodulated part of $\tilde{\eta}$ are determined by the experimental upper bounds on the unmodulated part of the rate with a procedure first outlined in Ref.~\cite{Fox:2010bz}. This procedure exploits the fact that, by definition, $\tilde{\eta}^0$ is a non-increasing function of $\vmin$. For this reason, the smallest possible $\tilde{\eta}^0(\vmin)$ function passing by a point $(v_0, \tilde{\eta}_0)$ in the $(\vmin, \tilde{\eta})$ plane is the downward step-function $\tilde{\eta}_0 \, \theta(v_0 - \vmin)$. That is, among the functions passing by the point $(v_0, \tilde{\eta}_0)$, the downward step-function is the one yielding the minimum predicted number of events. Imposing this functional form in \Eq{R1} we obtain
\beq
R_{[\Ed_1, \Ed_2]} = \tilde{\eta}_0 \int_0^{v_0} \ud \vmin \, \eR^{\rm SI}_{[\Ed_1, \Ed_2]}(\vmin) .
\eeq
The upper bound $R^{\rm lim}_{[\Ed_1, \Ed_2]}$ on the unmodulated rate is therefore translated into an upper bound $\tilde{\eta}^{\rm lim}(\vmin)$ on $\tilde{\eta}^0$ at $v_0$ by
\beq
\tilde{\eta}^{\rm lim}(v_0) = \frac{R^{\rm lim}_{[\Ed_1, \Ed_2]}}{\int_0^{v_0} \ud \vmin \, \eR^{\rm SI}_{[\Ed_1, \Ed_2]}(\vmin)} .
\eeq
The upper bound so obtained is conservative in the sense that any $\tilde{\eta}^0$ function extending even partially above $\tilde{\eta}^{\rm lim}$ is excluded, but not every $\tilde{\eta}^0$ function lying everywhere below $\tilde{\eta}^{\rm lim}$ is allowed \cite{Frandsen:2011gi}. We use this procedure to draw upper bounds from the SIMPLE, CDMS-II-Ge, CDMS-II-Si, CDMSlite, XENON10, XENON100 and LUX experiments.

In our halo-independent analysis the bounds from CDMS-II-Ge, CDMS-II-Si, CDMSlite, XENON10, XENON100 and LUX are derived as $90\%$ CL upper bounds using the Maximum Gap Method~\cite{Yellin:2002xd}. The SIMPLE bound is derived as the 90\% CL Poisson limit.
As for our SHM analysis, only sodium is considered for DAMA, and the quenching factor is taken to be $Q_{\rm Na} = 0.3$. For XENON10, limits produced by setting or not setting the electron yield $\mathcal{Q}_{\rm y}$ to zero below 1.4 keVnr (as in Ref.~\cite{Aprile:2012nq}) are obtained. For LUX, upper bounds considering 0, 1, 3 and 5 observed events are computed, as explained in Section~\ref{data}.

Figs.~\ref{eta7} and \ref{eta9} collect the results of our halo-independent analysis for a WIMP mass $m = 7$ GeV/$c^2$ and  $m = 9$ GeV/$c^2$, respectively: the left and right columns correspond to isospin-conserving ($f_n = f_p$) and isospin-violating ($f_n / f_p = -0.7$) interactions, respectively; and the top, middle and bottom rows show measurements and bounds for the unmodulated component $\tilde\eta^0 c^2$, for the modulated component $\tilde\eta^1 c^2$, and for both together, respectively, in units of day$^{-1}$. The middle row also shows the upper bounds on $\tilde\eta^0 c^2$ from the plots on the top row.

Although we show  up to $\vmin=$ 1000 km/s in our plots, $\vmin$ cannot be larger than the maximum WIMP speed in the halo, $v_{\rm max}$. Thus all the data points in Fig.~\ref{eta7} are allowed only if  $v_{\rm max} \gtrsim 950$ km/s, which are among the largest values for $v_{\rm max}$ found in the literature (see discussion after Eq.~\eqref{f_G}), while those in Fig.~\ref{eta9} require less extreme $v_{\rm max}$ values, $v_{\rm max} \gtrsim 750$ km/s.

 In the bottom panels of Figs.~\ref{eta7} and \ref{eta9} one can see that the CoGeNT 2014 modulation amplitude (solid line dark blue crosses) is  lower than the amplitude in the previous data (dashed blue crosses). It is a smaller fraction of the CoGeNT 2014 rate plus flat background (\ie without the flat and neutron backgrounds subtracted) than in the previous data. The CoGeNT 2014 rate is indicated with a solid horizontal brown line and the older rate with dashed horizontal brown line. Thus the tension between the relatively large modulation amplitude and the average rate plus background is less severe in the CoGeNT 2014 data than in the previous data. As already mentioned, we show the modulus of the negative modulation amplitude for the second bin with a thiner solid dark blue line.

The CDMSlite bound in the halo-independent analysis is much above the DM-signal regions, although it is the best available bound for isospin-violating interactions at $\vmin$ smaller than about 400 km/s (depending on the WIMP mass), and it is within a factor of $\sim 2$ of the CoGeNT unmodulated signal+background point around $\vmin \sim $ 450 km/s for a 7 GeV/$c^2$ mass, or 400 km/s for a 9 GeV/$c^2$ mass. At $\vmin \lesssim 600$ km/s for $m=7$ GeV/$c^2$, or 480 km/s for $m=9$ GeV/$c^2$, the most stringent bounds are from  XENON10 for the isospin-conserving case, and CDMS-II-Ge together with CDMS-II-Si for the isospin-violating case with $f_n / f_p = -0.7$.

The lesser impact of the CDMSlite bound in the halo-indendent analysis vs the SHM analysis stems from the steepness of the SHM $\tilde{\eta}^0(\vmin)$ curve as a function of $\vmin$. In the top left panel of Fig.~\ref{eta7} and the middle right panel of Fig.~\ref{eta9} we show the predicted $\tilde{\eta}^0 c^2$ (upper line) and $\tilde{\eta}^1 c^2$ (lower line) in the SHM for particular values of the WIMP-proton cross section. We choose these cross sections so that (a) for the $m = 7$ GeV/$c^2$ isospin-conserving case in the top left panel of Fig.~\ref{eta7}, the CDMS-II-Si unmodulated data are well explained by the SHM, \ie the $\tilde{\eta}^0$ curve passes through the red crosses in the figure (this happens for $\sigma_p = 10^{-40}$ cm$^2$), and (b)  for the $m = 9$ GeV/$c^2$ isospin-violating case in the middle right panel of Fig.~\ref{eta9},  the DAMA modulation data are well explained, \ie $\tilde{\eta}^1$ passes through the green crosses (this occurs for $\sigma_p = 2 \times 10^{-38}$ cm$^2$). These plots show that the $\tilde{\eta}^0(\vmin)$ of the SHM is a very steep function of $\vmin$ and thus can be constrained at low $\vmin$ values by the CDMSlite limit as well as by other upper limits on the unmodulated rate, as is reflected in Fig.~\ref{m-sigma}.

The LUX bound in the halo-independent analysis has a much weaker impact than it has in the SHM analysis. The conservative threshold of 3.0 keVnr on the scintillation/ionization signal does not allow the LUX bound to reach into the region with $\vmin \lesssim 600$ km/s (for $m=7$ GeV/$c^2$). Above $\vmin \sim 600$ km/s, the LUX bound is the most stringent bound, even in the isospin-violating case with $f_n / f_p = -0.7$.

For the isospin-conserving SI case, the DAMA and CoGeNT modulation points at $\vmin \gtrsim 600$ km/s for $m=7$ GeV/$c^2$, or at $\vmin \gtrsim 480$ km/s for $m=9$ GeV/$c^2$, which were in severe tension with the CDMS-II-Ge modulation bound, are now definitely incompatible with LUX. The isospin-conserving points at smaller $\vmin$ are in tension with the XENON10 bound, although the degree of disagreement depends on the choice of threshold for the electron yield $\mathcal{Q}_{\rm y}$. (Although the degree of disagreement cannot be readily quantified in our current halo-independent analysis, we expect that the deeper a point is inside the excluded region the higher the degree of disagreement between point and bound.)

For the isospin-violating case with $f_n / f_p = -0.7$, the DAMA and CoGeNT modulation points at $\vmin \lesssim 580$ km/s for $m=7$ GeV/$c^2$, or at $\vmin \lesssim 450$ km/s for $m=9$ GeV/$c^2$, are not excluded by other experiments in our halo-independent analysis. The other points are in strong tension with either the CDMS-II-Si DM-signal events, the CDMS-Ge modulation bound, or the LUX bound. Particularly puzzling is that the CDMS-II-Si DM-signal events (red crosses in Figs.~\ref{eta7} and \ref{eta9}) occur in this isospin-violating case at a rate smaller than the modulation amplitude suggested by the DAMA and CoGeNT data (green and blue crosses).

\section{Results and conclusions}

In the SHM analysis of the allowed regions and bounds in the $m$--$\sigma_p$ parameter space (Fig.~\ref{m-sigma}), CDMSlite and LUX set very stringent bounds, and together exclude the allowed regions of all four experiments with a positive signal (DAMA, CoGeNT, CRESST-II and CDMS-II-Si) for WIMPs with isospin-conserving couplings. The DAMA, CoGeNT 2014 and CRESST-II regions appear now to be largely excluded by the LUX limits also for WIMPs with isospin-violating spin-independent couplings ($f_n/f_p=-0.7$). In this case, also the XENON100 and CDMS-II-Si upper bounds significantly constrain the DAMA, CoGeNT 2014 and CRESST-II regions (other limits reject the DAMA region as well).

Although in our isospin-conserving SHM analysis the DM-signal region is severely constrained by the CDMSlite limit, in our halo-independent analysis this limit is much above the DM-signal region. The difference stems from the steepness of the SHM $\tilde{\eta}^0(\vmin)$ curve as a function of $\vmin$, which is constrained at low $\vmin$ by the CDMSlite (and other) limits. The conclusions of this analysis do not change when the uncertainty in the local value of the escape speed are taken into account, although the lowest mass reach of the LUX limit does change within a 25\% range.

In our halo-independent analysis, one should keep in mind that the data points presented in the plots are only allowed if they have a $\vmin$ value smaller than the maximum WIMP speed with respect to Earth in the local dark halo, which depends on the local escape speed with respect to the galaxy and  the velocity of Earth with respect to the galaxy. For $m$=7 GeV/$c^2$ all the data points considered are allowed only if  this maximum speed is larger than 950 km/s, namely among the largest values found in the literature, while those for $m=$ 9 GeV/$c^2$ require less extreme values, larger than 750 km/s. 

The new CoGeNT 2014 modulation amplitude (solid dark blue crosses) is smaller than the older modulation (dashed blue crosses), and this makes the modulation amplitude a smaller percentage of the average rate (plus background) than in the previous data. Fixing the modulation phase to be equal to that of DAMA's best fit and the period to one year, we find that the modulation amplitude  is compatible with zero at the $0.9\sigma$ and $1.1\sigma$ level, respectively, in the first and second energy bins we chose, and in the second energy bin it is negative.

In our halo-independent analysis we find that  although the LUX bound is more constraining than the XENON100 limit, both cover the same range in $\vmin$ space and are limited to $\vmin \gtrsim 600$ km/s for a WIMP mass of 7 GeV/$c^2$, or 500 km/s for 9 GeV/$c^2$. This is due to the conservative suppression of the response function below 3.0 keVnr. Thus the LUX bound and the previous XENON100 bound exclude the same data for isospin-conserving couplings. In other words, the DAMA, CoGeNT  and CDMS-II-Si energy bins that are not excluded by XENON100 are not excluded by LUX either. The situation, however, is of strong tension between the positive and negative results, as it was already before the LUX data.

LUX's improvement over XENON100 becomes important in our halo-independent analysis for isospin-violating couplings, which are chosen so that the DM interaction with Xe is highly suppressed. Together with the CDMS-II-Si bound, the LUX limit excludes or severely constrains the DAMA and CoGeNT 2014 modulation points, and the CRESST-II points, at $\vmin \gtrsim 600$ km/s for $m=7$ GeV/$c^2$, or $\vmin \gtrsim 500$ km/s for $m=9$ GeV/$c^2$. Still, as in our previous analysis~\cite{DelNobile:2013cta}, the most puzzling conundrum in the isospin-violating case with $f_n/f_p=-0.7$ is how the unmodulated rate as measured by CDMS-II-Si with its three events could be smaller than the modulation amplitude as measured by DAMA and CoGeNT.

\section*{Acknowledgments}

P.G. was supported in part by NSF grant PHY-1068111. E.D.N., G.G. and J.-H.H. were supported in part by Department
of Energy under Award Number DE-SC0009937. J.-H.H. was also partially supported by Spanish Consolider-Ingenio MultiDark (CSD2009-00064).

\end{document}